\documentclass[11pt,a4paper]{article}

\usepackage[a4paper]{geometry} 
\usepackage{titling} 
\usepackage{setspace} 
\usepackage{titlesec}

\usepackage[T1]{fontenc}
\usepackage{lmodern}
\usepackage[dvipsnames]{xcolor}
\usepackage{microtype}
\usepackage{dsfont}
\usepackage{parskip}

\usepackage{amsmath}
\usepackage{tabularx}
\usepackage{amssymb}
\usepackage{physics}
\usepackage{tensor}
\usepackage{braket}
\usepackage{slashed}
\usepackage{accents}

\usepackage{appendix}
\usepackage{comment}
\usepackage[colorlinks=true
  ,urlcolor=blue
  ,anchorcolor=blue
  ,citecolor=blue
  ,filecolor=blue
  ,linkcolor=blue
  ,menucolor=blue
  ,linktocpage=true
  ,pdfproducer=medialab
  ,pdfa=true
]{hyperref}
\usepackage{cleveref}

\usepackage{import}
\usepackage{standalone}

\usepackage{tikz} 
\usepackage{graphicx}
\usetikzlibrary{patterns}

\usepackage{array}   

\usepackage{cite}

\usepackage[width=.8\textwidth]{caption} 

\numberwithin{equation}{section}
\numberwithin{table}{section}
\begin{document}\spacing{1.2}
\vspace*{2.5cm}
\begin{center}
{\Large \textsc{Notes on complex $q=2$ SYK}}\\ 
\vspace*{1.3cm}
Ben Pethybridge\\ 
{
\footnotesize
Department of Mathematics, King's College London, Strand, London WC2R 2LS, UK \\
\footnotesize\textsf{ \href{mailto:ben.pethybridge@kcl.ac.uk}{ben.pethybridge@kcl.ac.uk}}
}
\end{center}
\vspace*{1.5cm}
\begin{abstract}
This note clarifies and extends results on the complex SYK model to the solvable $q = 2$ case. We calculate the four point function OPE of fermions in the low energy CFT, implying the existence of a tower of integer-weight operators in the IR. We comment on the lack of a mode breaking conformal symmetry in this special case of SYK and the consequences for deformations of the theory near the conformal fixed point. We use the nearly-free structure of the model to provide a closed form expression for OPE coefficients of the integer-weight operators. We also discuss analytic and numerical results relevant to the thermodynamics of $q=2$ SYK in both the complex and real case. The tower of operators transform in the discrete series of representations of $SL(2, \mathbb{R})$, the representations shared by dS$_2$ and AdS$_2$. In this work we continue discussion of holographic models including these representations \cite{Anninos:2023lin}. 
\noindent 
\end{abstract}
\newpage
\tableofcontents
\section{Introduction}\label{sec:intro}
We explore the proposal made in the outlook of  \cite{Anninos:2023lin}, that a highly constrained gauge theory in (A)dS$_2$ may have a microscopic description in terms of the $q=2$ SYK model. We examine the complexified version of the real SYK model introduced and described by Sachdev, Ye and Kitaev \cite{Sachdev:1993ppp,Kitaev:2015}. We summarise and extend the work of \cite{Maldacena:2016hyu,Gross:2017hcz,Bulycheva:2017uqj,Gu:2019jub} in the $q=2$ case and make new comments on the implications for conformal symmetry, the operator algebra and closed form expressions for OPE coefficients.

Our interest in this particular microscopic theory is the result of evolving developments in three connected areas of research. Firstly, the IR near-conformal limit of the generic $q>2$ case of the SYK model has attracted attention in recent years as a dual description of the black hole in near-AdS$_2$ JT gravity \cite{Jackiw:1984je,Teitelboim:1983ux}. In the $q > 2$ case there is an emergent reparameterisation symmetry which exhibits a pattern of symmetry breaking described in terms of the Schwarzian derivative effective action. This, along with the appearance of chaotic behaviour in the out-of-time-order correlators of the theory is indicative of the relationship to the 2 dimensional black hole \cite{Almheiri:2014cka,Kitaev:2015,Maldacena:2016upp,Cotler:2016fpe,Jensen:2016pah,Kitaev:2017awl,Joung:2023doq}, as reviewed in \cite{Sarosi:2017ykf,Rosenhaus:2018dtp}. This note performs analogous analysis to these references, and makes use of many of the results. However, we focus on the solvable $q=2$ case which is non-chaotic and therefore not regularly studied in the n-AdS$_2$ JT context. 

Instead the $q=2$ case has a potential relevance to a highly constrained, topological gauge theory in (A)dS$_2$ \cite{Anninos:2023lin}. The bulk theory is an $SL(N,\mathbb{R})$ BF theory in two dimensions, first explored in the context of low dimensional analogues to higher spin gravity in (A)dS by \cite{Alkalaev:2013fsa,Grumiller:2013swa}. This extended earlier work \cite{Bengtsson:1986zm,Vasiliev:1995sv} to write higher spin equations of motion for two-dimensions. This theory includes JT gravity as a subsector, coupled to matter in the form of a tower of integer-weight fields transforming in the discrete series representations of the (A)dS$_2$ isometry group $SO(1,2)$. Formal aspects of the higher spin algebra  and operator content of this theory were developed in \cite{Alkalaev:2020kut}. In parallel to the story of SYK in n-AdS JT, potential holographic duals have been discussed in AdS for the higher spin generalization of JT by \cite{Gonzalez:2018enk,Alkalaev:2019xuv}. This note aspires to be a part of this discussion, with a particular focus on the higher spin theory in dS$_2$ and reference to a  wider exploration of the physical consequences of de Sitter representation theory for quantum field theory \cite{Penedones:2023uqc,Baumann:2017jvh,Pethybridge:2021rwf,Schaub:2023scu,Letsios:2023qzq,Letsios:2024nmf,RiosFukelman:2023mgq}.

The third development relevant to this work is that of low dimensional models of quantum gravity in dS. Since initial discussions of holographic dS quantum gravity \cite{Witten:2001kn,Strominger:2001pn,Maldacena:2002vr}, there has been a concerted effort to find explicit models of dS/CFT. The proof of principal case is the description of Vasiliev higher spin gravity  in dS$_4$ by free bosonic degrees of freedom living at the late time boundary  \cite{Anninos:2011ui,Anninos:2017eib}, for which our proposed description may be thought of as a lower dimensional analogue. We are motivated to find holographic descriptions of discrete series models in particular, as these are the representations, in two- and four-dimensions, which are carried by the graviton in dS and AdS. The common existence of discrete series models in AdS$_2$ and dS$_2$ also allows for a more direct comparison, making these important tools in building models of dS/CFT. There have been a number of alternate research avenues opened in recent years \cite{Martinec:2014uva,Bautista:2015wqy,Bautista:2019jau,Anninos:2021ene,Muhlmann:2021clm,Coleman:2021nor,Muhlmann:2022duj}, some of which attempt to find low-dimensional toy models which carry some of the properties of four-dimensional de Sitter quantum gravity. Examples which share some of the areas of interest in our work include interpretations of JT gravity in de Sitter \cite{Maldacena:2019cbz,Cotler:2019nbi, Svesko:2022txo, Nanda:2023wne, Anninos:2022hqo}, and considerations of embedding a piece of dS$_2$ in AdS \cite{Anninos:2017hhn,Anninos:2018svg,Anninos:2022qgy,Anninos:2024wpy}. Within this area there has also been a recent interest in a potential connection between the  Double-scaled SYK model (DSSYK) and dS$_3$ \cite{Susskind:2021esx,Rahman:2022jsf,Rahman:2024vyg,Narovlansky:2023lfz,A:2023psv,Verlinde:2024znh}, a separate but complimentary conjecture to that of \cite{Anninos:2023lin} and this note. 

This paper is structured as follows. We first define the quantum mechanical complex $q=2$ SYK model in \cref{sec:model}, describing the classical equations of motion, and rewriting the action in terms of the bilinear ``master fields'' $(G,\Sigma)$. We also calculate the two point function from the Schwinger Dyson equations of motion in the large $N$ limit. We consider correlation functions of the large $N$ IR CFT in \cref{sec:Corfunctions}. In this section we make the claim that there exist a tower of integer-weight operators in the CFT which contribute as conformal blocks to the four point function of fermions, we contrast this with the case of general $q$ in which the operators pick up non-integer weights, and comment on consequences for conformal symmetry near the IR fixed point. We also construct OPE coefficients for these operators in terms of a finite series. Finally in the outlook \cref{sec:out} we comment on the thermodynamics of the model, using numerical results presented in \cref{App:thermo} and analytic arguments to comment on differences with the general $q$ model. 
\section{Complex \texorpdfstring{$q=2$}{q=2} SYK model}\label{sec:model}
We focus on the complex SYK model in the special case of a quadratic interaction. This model is the same as that of \cite{Gu:2019jub,Bulycheva:2017uqj, Afshar:2019axx} with the interaction  parameter $q = 2$. We begin with a classical analysis of the equations of motion and a short discussion of quantization. Following the usual approach for $q>2$ \cite{Maldacena:2016hyu}, we write the model as an effective theory of a pair of bilinear operators (the $(G,\Sigma)$ formalism). 

We consider a theory of $N$ complex fermions in a single Euclidean dimension parameterised by $\vartheta$. To transform between the line and the thermal circle we use the conformal transformation
\begin{equation}
    \vartheta_{\text{line}}  = \tan{\frac{\pi }{\beta}}\vartheta_\text{circle}~,
\end{equation}
where $\beta$ is the periodicity of the thermal circle. The complex field $\psi_i$ is built out of a pair of one-dimensional Grassman-odd Majorana fermions  $\overline{\chi}_i^I = \chi_i^I$, such that 
\begin{align}
    \psi_i = & \chi_i^1 + i \chi_i^2~,\\
    \overline{\psi}_i = & \chi_i^1 - i \chi_i^2~.
\end{align}
We keep the convention that for any spinor, real or complex $ \overline{(\xi\eta)} = \overline{\eta}\overline{\xi}$. The Dirac spinor field is also Grassman odd, as a result $(\psi_i)^2 =0$ and $\overline{\psi}_i\psi_i =-\psi_i  \overline{\psi}_i \neq 0$. The Hamiltonian is

\begin{equation}
    H_{SYK} =\frac{i}{2} J_{ij}\overline{\psi}_i\psi_j ~,
\end{equation}
where $J_{ij}$ is an antisymmetric, real matrix to preserve hermiticity of the Hamiltonian and repeated indices are summed over. The $J_{ij}$ are drawn from a Gaussian ensemble with variance $ \frac{J^2}{N}$. The theory is the result of taking an average over these quadratic ``realisations'' of the model. 

The conjugate momentum of $\psi_i$ is $ \overline{\psi_i}$ and the Hamiltonian equations of motion are
\begin{align}
    \partial_\vartheta\psi_i &= \frac{i}{2}J_{ij}\psi_j~,\\
    \partial_\vartheta\overline{\psi}_i &= -\frac{i}{2}J_{ij}\overline{\psi}_j~.
\end{align}
Taking the Legendre transform we have the Euclidean action
\begin{equation*}
    S_{SYK}=\frac{1}{2}\int d\vartheta\left[\Bar{\psi}_i\partial_\vartheta{\psi}_i-i J_{ij} \Bar{\psi}_i\psi_j\right]~.
\end{equation*}
For a particular realisation of the model, with fixed $J_{ij}$, the equation of motion may be solved directly by diagonalising the matrix $J_{ij}$. As such, a single realisation is a model of $N$ free fermions in one dimension with random masses, the only source of subtlety is the average over the $J_{ij}$, referred to here as the disorder average. The disorder average will either be performed as a ``quenched'' average, in which case the partition function will be averaged and thermodynamic quantities calculated from this partition function. Or, in the case of a few thermodynamic quantities in the outlook, an ``annealed'' average where the calculation will be performed for the free random mass fermion and then averaged. As in the general $q$ case we expect the two quantities to converge in the large $N$ limit. In this note annealed average quantities will be denoted with angular brackets $\braket{\mathcal{O}}_J$, quenched averaged quantities will not. 

The almost free nature of the theory is reflected in the enhanced symmetry at the classical and quantum level, as discussed in the outlook of  \cite{Anninos:2023lin}. The symmetries of this model will be further discussed in the IR CFT limit below. It is expected, and a result of this paper that there should be a tower of operators in the IR spectrum of the theory. There is already a suggestion of this in the fact that the underlying theory is free, and therefore has an number of conserved charges. In addition to the $ U(1) $ charge 
\begin{equation}
    \mathcal{O}_1  = \overline{\psi}_i\psi_i~,
\end{equation}
and the $h =2$ mode of \cite{Maldacena:2016hyu} (here generated  classically by the global spacetime symmetry $ \psi_i \rightarrow \psi _ i + \epsilon \partial_\vartheta \psi_i$)   
\begin{equation}
    \mathcal{O}_2  = \overline{\psi}_i\partial_\vartheta\psi_i~,
\end{equation}
there are further conserved charges, for example 
\begin{equation}
    \mathcal{Q} =   J_{ij} J_{jk}\overline{\psi}_i\psi_k~.
\end{equation}
These are only the local transformations, if we enlarge the set to non-local transformations of the fields we can find an infinite class of symmetries for the model, even away from the IR fixed point (see equation (7.20) in \cite{Anninos:2023lin}). 

On quantization we apply the canonical anti-commutation relation 
\begin{equation}
    \{ \overline{\psi}_i, \psi_j\} =\delta_{ij}~.
\end{equation}
Considering the vacuum $\ket{0}$ annihilated by $\psi_i$ and, using $(\overline{\psi}_i)^2 =0$, we build a finite basis of n-particle states 
\begin{equation}
\ket{i_\alpha;n} = \prod_{\alpha = 1}^n\overline{\psi}_i\ket{0}~.
\end{equation}
The states are characterised by the number of particles, $ 0 \geq n \geq N$, the eigenvalue of the number operator $\overline{\psi}_i  \psi_i$. The number of possible states at particle number $n$ is $\binom{N}{n}$, implying a $2^N$-dimensional Hilbert space.  
\subsection{\texorpdfstring{$(G,\Sigma)$}{(G ,S)} formalism and two-point function} 
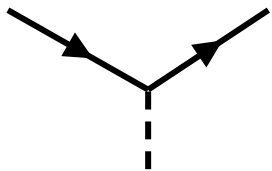
\begin{figure}[h!]
    \centering
    \tikzset{every picture/.style={line width=0.75pt}} 

\begin{tikzpicture}[x=0.75pt,y=0.75pt,yscale=-1,xscale=1]

\draw [line width=2.25]    (130,10) -- (200,50) ;
\draw [shift={(172.47,34.27)}, rotate = 209.74] [fill={rgb, 255:red, 0; green, 0; blue, 0 }  ][line width=0.08]  [draw opacity=0] (14.29,-6.86) -- (0,0) -- (14.29,6.86) -- cycle    ;
\draw [line width=2.25]    (200,50) -- (260,10) ;
\draw [shift={(237.16,25.23)}, rotate = 146.31] [fill={rgb, 255:red, 0; green, 0; blue, 0 }  ][line width=0.08]  [draw opacity=0] (14.29,-6.86) -- (0,0) -- (14.29,6.86) -- cycle    ;
\draw [line width=2.25]  [dash pattern={on 6.75pt off 4.5pt}]  (200,90) -- (200,50) ;

\end{tikzpicture}
    \caption{The effective interaction vertex of free fermions with a disorder line.}
    \label{fig:int}
\end{figure}

The quenched disorder average is performed by allowing the $J_{ij}$ to vary in the path integral
\begin{equation}\label{def:fullZ}
    Z =\braket{\mathcal {Z}}_J = \int[DJ_{ij}][D\psi_i][D\overline{\psi}_i]e^{- \frac{\sqrt{N} J_{ij}J_{ij}}{2J}}e^{-S_{SYK}} ~.
\end{equation}
This model is not quite quadratic, in the sense that it includes a cubic ``interaction'' term resulting in Feynman diagrams (\cref{fig:int}) involving an effective vertex from the disorder average, which does not transfer momentum.
A finite $N$ analysis of certain observables may be performed by rewriting the path integral over $J_{ij}$ in terms of its eigenvalues, this approach is considered in appendix 2 of \cite{Gross:2016kjj}. Here we work in the limit of large $N$ to simplify the analysis of the two- and four-point function. 

At infinite $N$ the only contributing diagrams to the two-point function are non-crossing ``rainbow'' diagrams, the $q = 2$ version of the ``melonic'' diagrams for ordinary SYK \cite{Gross:2016kjj,Sarosi:2017ykf}. One may perform the following analysis using these diagrams to gain the large $N$ formula for the two-point function $G(\vartheta,\vartheta') = \braket{\overline{\psi}_i(\vartheta)\psi_i(\vartheta')}$, where the repeated index is summed over. Alternatively, one can start from the path integral \cref{def:fullZ} and perform the integral over $J_{ij}$, which appears quadratically. By integrating in the identity in terms of $G$
\begin{equation}
    G(\vartheta,\vartheta')  = \frac{\braket{\overline{\psi}_i(\vartheta) \psi_i (\vartheta')}}{N} \implies 1 = \int [DG] \delta(N G-\overline{\psi}_i(\vartheta) \psi_i (\vartheta')) ~,
\end{equation}
we perform the path integral over $\psi_i$ to give the final result 
\begin{equation}
    Z = \left(\frac{N i \pi}{J^2}\right)^{\frac{N(N+1)}{2}}\int [DG][D\Sigma] e^{-N I[G,\Sigma]}~,
\end{equation}
such that
\begin{equation}\label{def:Effact}
    I[G,\Sigma]  = \frac{1}{2}\left(-\log[\det(\partial_\vartheta -\Sigma)]+\int d\vartheta d\vartheta' \left[ \Sigma(\vartheta-\vartheta')G(\vartheta-\vartheta')-\frac{J^2}{2} G(\vartheta-\vartheta')G(\vartheta-\vartheta')\right]\right)~.
\end{equation}
The equations of motion for the above action are the Schwinger-Dyson equations of q = 2 SYK:
\begin{align}
\partial_\vartheta G(\vartheta,\vartheta') -\int d\upsilon \Sigma(\vartheta,&\upsilon)G(\upsilon,\vartheta) = \delta(\vartheta-\vartheta')~,\label{eq:eom1}\\
    \Sigma (\vartheta ,\vartheta')& = J^2 G(\vartheta,\vartheta') ~.
\end{align}
We are permitted to perform a semi-classical analysis due to the appearance of $N$ multiplying the action $I[G, \Sigma]$. We solve the Schwinger Dyson equations at finite cutoff, as well as considering the UV and IR limits.
\paragraph{Frequency space:}
Examining the Schwinger Dyson equations in frequency space is simplest and provides a single quadratic equation of motion 
\begin{align}
    i \omega G(\omega )-J ^2 G(\omega) ^2 = 1 ~,
\end{align}
the solution to which is
\begin{equation} \label{eq:Ggen}
    G(\omega) = \frac{i\text{ sgn}(\omega)}{2J^2}\left(\abs{\omega}- \sqrt{\omega^2 + 4 J^2}\right) ~.
\end{equation}
Observing that $J$ scales with energy, implying the UV exists at high frequency relative to $J$, $G(\omega)$  in this case cannot be obtained from \cref{eq:Ggen}. Instead, solving the Schwinger Dyson equations once more gives
\begin{equation}
    G_{UV}(\omega) = -\frac{i}{\omega}~,
\end{equation}
while in the IR ($J\rightarrow \infty$) 
\begin{equation} \label{eq:freIRG}
    G_{IR}(\omega) = -\frac{i \text{ sgn}(\omega)}{J}~.
\end{equation} 
\paragraph{Position Space:}
In the case of free fermions the Hamiltonian is taken to be zero as the kinetic term dominates the action, this implies the field itself does not evolve in time $\psi_i(\vartheta) \approx \psi_i(0)$, we may them use the commutation relation, along with translation invariance and antisymmetry to find 
\begin{equation}
    G_{UV}(\vartheta) = \frac{ \text{sgn}(\vartheta)}{2}~. 
\end{equation}
In the IR limit we would expect the effective description in terms of G to take the form of a CFT two-point function, using the ansatz
\begin{equation}
    G_{IR}(\vartheta) = \frac{b}{|\vartheta|^{2\Delta}}~, 
\end{equation}
and substituting this into the fourier transform for the IR two point function in frequency space demands 
\begin{equation} \label{def:GCFT}
    G_{IR}(\vartheta) = \frac{1}{J \pi  \vartheta} ~. 
\end{equation}
This implies the existence of an operator in the IR CFT which we continue to label $\psi_i$, which transforms as a $\Delta = \tfrac{1}{2}$ primary. In addition the full position space two point function is accessible in this solvable case, taking the fourier transformation of \cref{eq:Ggen}
\begin{equation}
    G(\vartheta) = \frac{I_1(2 J | \vartheta | )-\pmb{L}_1(2 J |\vartheta|
   )}{2 J |\vartheta|}~.
\end{equation}
Where the $I_1$ and $\pmb{L}_1$ are spherical Bessel and Struve special functions respectively. We note that we can take the  large and small $J \vartheta$ limit to find the UV and IR formulae respectively.
\section{Correlation functions in the large \texorpdfstring{$N$}{N} limit}\label{sec:Corfunctions}
We now turn to the decomposition of the four point function of fermions in terms of primaries in the IR CFT. Our objective is to show the existence of a tower of integer-weight operators in the spectrum of the CFT. We then consider the OPE coefficients of these operators. 
\subsection{Four-point function}
Here we follow the analysis of \cite{Bulycheva:2017uqj}, specialising the results of this work to the solvable case of $q = 2$. The object of interest is the four-point function
\begin{equation}
\frac{\braket{\bar{\psi}_i(\vartheta_1)\psi_i(\vartheta_2)\bar{\psi}_j(\vartheta_3)\psi_j(\vartheta_4)}}{N^2}~.
\end{equation}
Schematically the leading and subleading terms of the four point function are 
\begin{equation}
G(\vartheta_1-\vartheta_2) G(\vartheta_3-\vartheta_4)+\frac{1}{N} \sum_{n=0}^\infty \mathcal{F}_n(\vartheta_1,\vartheta_2,\vartheta_3,\vartheta_4)+ \mathcal{O}\left(\frac{1}{N^2}\right)~.
\end{equation}
Where the sum over $n$ expands the four point function in terms of ladder diagrams with $n$ ``rungs'' constructed of disorder lines. $\mathcal{F}_n$ is organised using symmetry under the operation of swapping the fermions. Assuming the four point function is even under time-reversal there may be only two contributions, using the notation of \cite{Bulycheva:2017uqj}:
\begin{equation}
\mathcal{F}_n \equiv \mathcal{F}_n^+ -\mathcal{F}_n^-~, 
\end{equation}
where $\mathcal{F}_n^+$ and $\mathcal{F}_n^-$  are respectively even and odd under operations which act as either $\vartheta_1 \leftrightarrow\vartheta_2$ or $\vartheta_3 \leftrightarrow \vartheta_4$. Each is the contribution of a ladder diagram which may be constructed by repeated integration against a kernel $K$, 
\begin{equation}\label{def:kernelac}
    \mathcal{F}^\pm_n(\vartheta_1,\vartheta_2,\vartheta_3,\vartheta_4) = \int d\vartheta d\vartheta' K(\vartheta_1 ,\vartheta_2;\vartheta,\vartheta')\mathcal{F}^\pm_{n-1} (\vartheta, \vartheta',\vartheta_3,\vartheta_4)~, 
\end{equation}
where 
\begin{equation}\label{def:kern}
    K(\vartheta_1,\vartheta_2;\vartheta,\vartheta') \equiv-J^2 G(\vartheta_1-\vartheta) G(\vartheta_2-\vartheta')~. 
\end{equation}
The initial contribution, without any disorder connecting the two branches is 
\begin{equation}
     \mathcal{F}^\pm_0(\vartheta_1, \vartheta_2,\vartheta_3,\vartheta_4) = G(\vartheta_1-\vartheta_3)G(\vartheta_2-\vartheta_4)\pm G(\vartheta_1-\vartheta_4)G(\vartheta_2-\vartheta_3)~.
 \end{equation}
Schematically, for both symmetric and antisymmetric contributions we resum the ladder diagrams using the following formula, as described in \cite{Sarosi:2017ykf,Maldacena:2016hyu, Gross:2016kjj,Bulycheva:2017uqj} for the general $q$ model,
\begin{equation}\label{def:hdecmp}
      \sum_{n=0}^\infty \mathcal{F}^\pm_n = \sum_h  \frac{1}{1-k(h)}\frac{\braket{\Psi^\pm_h, \mathcal{F}^\pm_0}}{\braket{\Psi_h^\pm,\Psi_h^\pm}}\Psi^\pm_h~.
\end{equation}
Where the sum includes a contribution for all non-zero eigenvalues, $k(h)$, of the operation of the kernel defined in \cref{def:kernelac}. The eigenfunctions of the kernel, $\Psi_h$, for symmetric and antisymmetric contributions were calculated in \cite{Bulycheva:2017uqj}; their conformally invariant form, in terms of the cross ratio is included in \cref{app:eigen}. Here we have been careful to take the large-$N$ limit and resum the series of ladder diagrams (i.e. sum over $n$) before taking the CFT limit. Failure to use this order of limits results in ambiguities, indeed we must have conformal symmetry to make precise the meaning of $h$, which is the eigenvalue of the conformal Casimir.

An important simplification can be made in the case of the $q = 2$ model. None of the ladders have any momentum transfer between the branches, as described for the real $q = 2 $ SYK model in appendix C of \cite{Gross:2016kjj}. In the above formalism, this fact is reflected in the independence of $k^\pm(h)$ from $h$,
\begin{equation}
    k^\pm(h) = -1~.
\end{equation}
This can be directly calculated from the three point function of operators of weight $h$ with two fermions, as shown in \cref{app:evals}. We immediately write the result 
\begin{equation}\label{eq:invariant}
     \sum_{n=0}^\infty \mathcal{F}^\pm_n = \frac{1}{2}\sum_h \frac{\braket{\Psi^\pm_h, \mathcal{F}^\pm_0}}{\braket{\Psi_h^\pm,\Psi_h^\pm}}\Psi^\pm_h = \frac{1}{2} \mathcal{F}^\pm_0~.
\end{equation}
Therefore 
\begin{equation} \label{eq:wick4pt}
\frac{\braket{\bar{\psi}_i(\vartheta_1)\psi_i(\vartheta_2)\bar{\psi}_j(\vartheta_3)\psi_j(\vartheta_4)}}{N^2}  = G(\vartheta_1-\vartheta_2) G(\vartheta_3-\vartheta_4)+\frac{1}{N} G(\vartheta_1-\vartheta_4) G(\vartheta_2-\vartheta_3)+ \mathcal{O}\left(\frac{1}{N^2}\right)~.
\end{equation}
\subsection{Harmonic analysis}
The benefit of writing the four point function as a sum over ladder contributions is the ease with which we may now decompose the four point function into contributions of conformal primaries, as noted first in \cite{Maldacena:2016hyu}. In this section we perform a harmonic analysis to prove the existence of a tower of integer-weight operators in the spectrum of the theory. The invariant cross ratio in one dimension is 
\begin{equation}
    x = \frac{\vartheta_{12}\vartheta_{34}}{\vartheta_{13}\vartheta_{24}} \hspace{1cm},\hspace{1cm}  1-x = \frac{\vartheta_{23}\vartheta_{14}}{\vartheta_{13}\vartheta_{24}}~,
\end{equation}
where $\vartheta _{ij} = \vartheta _i-\vartheta_j$. Considering the conformally invariant four point function
\begin{equation}\label{eq:q24p}
   \frac{\braket{\bar{\psi}_i(\vartheta_1)\psi_i(\vartheta_2)\bar{\psi}_j(\vartheta_3)\psi_j(\vartheta_4)}}{N^2\braket{\overline{\psi}_i(\vartheta_1)\psi_i(\vartheta_2)}\braket{\overline{\psi}_i(\vartheta_3)\psi_i(\vartheta_4)}} = 1+ \frac{1}{N}\left( \frac{x}{1-x}\right)+ \mathcal{O}\left(\frac{1}{N^2}\right)~,
\end{equation}
here we have used $ 1-x = \frac{\vartheta_{23}\vartheta_{14}}{\vartheta_{13}\vartheta_{24}}$. In another form the four point function  may be written as a sum over conformal blocks,
\begin{equation}\label{def:4ptOPE}
     1 + \left(\sum_h(c_{\psi\psi}^h)^2\tilde{\Psi}_h(x)\right) \equiv 1+ \frac{1}{N}\left( \frac{x}{1-x}\right)+ \mathcal{O}\left(\frac{1}{N^2}\right)~. 
\end{equation}
Where $h$ are the eigenvalues of the conformal Casimir,
\begin{equation}
    \mathcal{C} = x^2(1-x)\partial^2_x-x^2\partial_x~.
\end{equation}
The functions $\tilde{\Psi}_h(x)$ are the eigenfunctions. As the conformal Casimir and  the action of the kernel defined in \cref{def:kernelac} commute, these are precisely the functions used to decompose the four point function in terms of ladder diagrams in \cref{def:hdecmp}, made conformally invariant.
\begin{equation}
    \tilde{\Psi}_h(x) \equiv \frac{\Psi_h(\vartheta_1,\vartheta_2,\vartheta_3,\vartheta_4)}{\braket{\overline{\psi}_i(\vartheta_1)\psi_i(\vartheta_2)}\braket{\overline{\psi}_i(\vartheta_3)\psi_i(\vartheta_4)}}~,
\end{equation}
We use $\sim$ to describe the conformally invariant versions of the functions above in the following. Comparing \cref{def:4ptOPE} to \cref{eq:invariant} we gain the following formula for the OPE coefficients 
\begin{equation}
    (c^{\pm,h}_{\psi\psi})^2 = \pm\frac{\braket{\braket{\tilde{\Psi}^\pm_h,\tilde{\mathcal{F}}_0^\pm}}}{2N\braket{\braket{\tilde{\Psi}^\pm_h,\tilde{\Psi}^\pm_h}}}+\mathcal{O}\left(\frac{1}{N^2}\right)~.
\end{equation}
In \cref{app:eigen} there are formulae for $\braket{\tilde{\Psi}^\pm_h,\tilde{\Psi}^\pm_{h'}}$, the only calculation that remains is to find $\braket{\tilde{\Psi}^\pm_h,\tilde{\mathcal{F}}_0^\pm}$, in terms of the cross ratio 
\begin{equation}\label{def:F0x}
    \tilde{\mathcal{F}}_0^\pm = \frac{1}{2}\left(x \pm \frac{x}{x-1}\right)~.
\end{equation}
The inner product in the space of eigenfunctions is 
\begin{equation}
\braket{\braket{f,g}} \equiv \int_0^\infty \frac{dx}{x^2}\bar{f}g~. 
\end{equation} 
Making use of the symmetry of \cref{def:F0x} under $x\rightarrow \frac{x}{x-1}$ , these are calculated exactly in equation (A.11) of \cite{Bulycheva:2017uqj} 
using the split representation, and agree  with \cite{Maldacena:2016hyu} for the symmetric pieces, specialising to the $q = 2$ case
 \begin{equation}
     \braket{\braket{\tilde{\Psi}^\pm_h,\tilde{\mathcal{F}}_0^\pm}} = \int\frac{dx}{x}\tilde{\Psi}_h^\pm(x) = -\pi^2~.
 \end{equation}
Generally, the four point function may be written with contributions split between the principal series $h = \frac{1}{2}+is$  and the discrete series $h \in \mathbb{Z}^+$, as written explicitly in equation $(5.42)$ and $(5.43)$ of \cite{Bulycheva:2017uqj}. We note that the wavefunctions $\tilde{\Psi}^\pm_h$ are symmetric under the ``shadow'' transformation $ h\rightarrow 1-h$. It is well known from \cite{Maldacena:2016hyu}, that for $q>2$ the symmetric contribution may be rewritten as a single sum over a discrete set of poles with irrational weights. We now discuss the analogous construction for the $q =2$ model. Combining these results 
\begin{equation}
    \mathcal{F}(x) = \frac{1}{4}\int^{\tfrac{1}{2} +i \infty}_{\tfrac{1}{2}-i\infty} dh \frac{2h -1}{\tan\pi h} \left(\tilde{\Psi}_{h}^+(x)-\tilde{\Psi}_{1-h}^-(x)\right) + \sum_{n=1}^\infty \frac{2n-1}{4}  \tilde{\Psi}(x).
\end{equation}
Where we have defined $ \tilde{\Psi}_n(x) = \pm\tilde{\Psi}^\pm_n(x)$ for respectively even, and odd values of $n$. 

For the principal series contribution, we have used the shadow symmetry $s \rightarrow -s$ to define the contour along the entire real axis in s. Using the appendix, in analogy to equation (3.84) in \cite{Maldacena:2016hyu} we can write the contour integral over the principal series in terms of the hypergeometric functions
\begin{equation}
     \int^{\tfrac{1}{2} +i \infty}_{\tfrac{1}{2}-i\infty} dh \frac{2h -1}{\tan\pi h} \chi^P_s(x) ~,
\end{equation}
where we can rewrite the wavefunctions using a hypergeometric identity such that
\begin{equation}
  \chi^P_s(x) = \begin{cases} \frac{\Gamma(h)^2}{\Gamma(2h)} x^h \, _2F_1(h,h;2 h;x) \hspace{1cm} x<1~,\\ \frac{\pi}{\sin \pi h}\, _2F_1(h,1-h;1;\frac{1}{x}) \hspace{1cm} x>1~.\end{cases}
\end{equation}
In both cases there are poles at integer points in $h$, which we pick up by deforming the contour onto the positive real axis, the result is 
\begin{equation}\label{eq:4ptsum}
    \mathcal{F}(x) =\sum_{n = 1}^\infty \frac{(2n-1)}{2} \Psi_n(x)~.
\end{equation}
Where $\Psi_n$ are the discrete series eigenfunctions of the conformal casimir, as defined in \cref{app:eigen}. It is possible to verify this relationship numerically by resumming a finite number of the hypergeometrics to recover \cref{eq:4ptsum}.

This is the most important result of this note, and has not featured prominently in previous literature. There is a significant difference in the $q=2$ case displayed here from the generic $q>2$ case of \cite{Maldacena:2016hyu,Gross:2016kjj,Bulycheva:2017uqj}. In the general model, the contour deformation replaces the discrete and principal series expansion by a sum over operators with irrational weights, here they have exact integer weights. The existence of a tower of operators transforming in a representation of $SL(2, \mathbb{R})$ with discrete series weights in the $q=2$ model links this theory to the models discussed in \cite{Anninos:2023lin}. In addition, the exact integer-weights are fundamentally important to dS$_2$. Usually in AdS the timelike direction is decompactified by taking the universal cover of the $SO(2,1)$ isometry group. This group has a more general set of highest or lowest weight representations with weights in $\mathbb{R}^+$. In de Sitter no such universal cover is taken, and so highest or lowest weight representations have protected integer weights.  

Additionally, a very important contribution to the four point function in the real $q>2$ case is the double pole at $h=2$ \cite{Maldacena:2016upp, Kitaev:2017awl, Sarosi:2017ykf}. In the complex model there is an infinite term in the sum at either $n = 1,2$, as described in the generic $q$ case for the complex model in \cite{Bulycheva:2017uqj}. These modes break the conformal reparameterisation invariance of the general model, leading to the domination of the leading order correction to the conformal action (the Schwarzian action) in the soft sector of the model and it's description as a ``near-CFT'' in the literature. The $q=2$ model seems to sidestep this complication, instead we anticipate that moving away from the IR fixed point will involve a more equal contribution from the tower of operators derived in \cref{eq:4ptsum}. The Schwarzian action will still make an appearance, however we anticipate additional structure in the form of ``higher spin'' versions of the Schwarzian similar to those discussed in \cite{Gonzalez:2018enk}. 
\subsection{OPE coefficients}
Looking toward future discussion of the bulk theory, we now write a closed formula for the OPE coefficients $c_{nmk}$ appearing in the three point function of the primary integer-weight operators to leading non-trivial order in $\frac{1}{N}$.
\begin{equation}
\label{def:cft3pt} \braket{\mathcal{O}_n(\vartheta_1)\mathcal{O}_m(\vartheta_2)\mathcal{O}_k(\vartheta_3)} = \frac{1}{\sqrt{N}}\frac{c_{nmk}}{|\vartheta_{12}|^{n+m-k}|\vartheta_{13}|^{n+k-m}|\vartheta_{32}|^{m+k-n}}~,
\end{equation}
This calculation in the generic $q>2$ case has been performed explicitly in \cite{Gross:2017hcz}, the formulae included simplify considerably for $q=2$; some of their considerations are unnecessary due to the almost free nature of the theory. 
\subsubsection{Fermion six-point function}
We first write a closed form expression for the connected part of the six point function of fermions, to leading order in $\frac{1}{N}$ \cite{Gross:2017hcz}. In this section ``connected'' is used in the sense that all propagators are connected by disorder lines. However, in the final result the connected piece will still be a reorganised product of propagators. This combinatorial exercise is accomplished by considering combinations of four point functions.

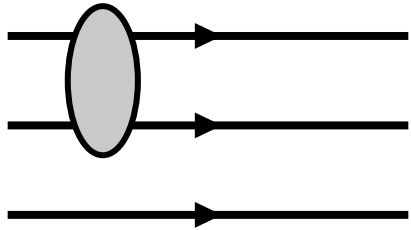
\begin{figure}[h!]
    \centering
    \tikzset{every picture/.style={line width=0.75pt}} 

\begin{tikzpicture}[x=0.75pt,y=0.75pt,yscale=-0.5,xscale=0.5]

\draw [line width=3]    (0,230) -- (400,230) ;
\draw [shift={(214,230)}, rotate = 180] [fill={rgb, 255:red, 0; green, 0; blue, 0 }  ][line width=0.08]  [draw opacity=0] (27.15,-13.04) -- (0,0) -- (27.15,13.04) -- cycle    ;
\draw [line width=3]    (0,140) -- (400,140) ;
\draw [shift={(214,140)}, rotate = 180] [fill={rgb, 255:red, 0; green, 0; blue, 0 }  ][line width=0.08]  [draw opacity=0] (27.15,-13.04) -- (0,0) -- (27.15,13.04) -- cycle    ;
\draw [line width=3]    (0,50) -- (400,50) ;
\draw [shift={(214,50)}, rotate = 180] [fill={rgb, 255:red, 0; green, 0; blue, 0 }  ][line width=0.08]  [draw opacity=0] (27.15,-13.04) -- (0,0) -- (27.15,13.04) -- cycle    ;
\draw  [color={rgb, 255:red, 0; green, 0; blue, 0 }  ,draw opacity=1 ][fill={rgb, 255:red, 201; green, 201; blue, 201 }  ,fill opacity=1 ][line width=2.25]  (60,95) .. controls (60,53.58) and (75.67,20) .. (95,20) .. controls (114.33,20) and (130,53.58) .. (130,95) .. controls (130,136.42) and (114.33,170) .. (95,170) .. controls (75.67,170) and (60,136.42) .. (60,95) -- cycle ;

\end{tikzpicture}
    \caption{Diagrams of this form contribute to the six point function of fermions at first order in $\frac{1}{N}$. However they only contribute as a contact term to the three point function of integer-weight operators. }
    \label{fig:sixp}
\end{figure}

In an individual realisation of the model, disconnected diagrams occur at every order in $N$, they are represented by the type of diagram displayed in \cref{fig:sixp}, where the grey shaded area is a sum over the entire expansion of the four point function in all powers of $\frac{1}{N}$. These disconnected diagrams do not contribute to the three point function of the integer-weight operators except as contact terms, as a consequence of the particular coincident point limits we take below, here we ignore them. 

The important contribution comes from connected diagrams. The first leading-order connected contribution is a sum over two sets of ladder diagrams which sequentially occur between two of the propagators, at order $N^{-2}$.  
\begin{figure}[h!]
    \centering
    \tikzset{every picture/.style={line width=0.75pt}} 

\begin{tikzpicture}[x=0.75pt,y=0.75pt,yscale=-1,xscale=1]

\draw [line width=3]    (0,118.51) -- (170,118.51) ;
\draw [shift={(94.2,118.51)}, rotate = 180] [fill={rgb, 255:red, 0; green, 0; blue, 0 }  ][line width=0.08]  [draw opacity=0] (13.57,-6.52) -- (0,0) -- (13.57,6.52) -- cycle    ;
\draw [line width=3]    (0,74.85) -- (170,74.85) ;
\draw [shift={(94.2,74.85)}, rotate = 180] [fill={rgb, 255:red, 0; green, 0; blue, 0 }  ][line width=0.08]  [draw opacity=0] (13.57,-6.52) -- (0,0) -- (13.57,6.52) -- cycle    ;
\draw [line width=3]    (0,31) -- (170,31) ;
\draw [shift={(94.2,31)}, rotate = 180] [fill={rgb, 255:red, 0; green, 0; blue, 0 }  ][line width=0.08]  [draw opacity=0] (13.57,-6.52) -- (0,0) -- (13.57,6.52) -- cycle    ;
\draw [line width=3]  [dash pattern={on 7.88pt off 4.5pt}]  (29.75,31.19) -- (29.75,74.85) ;
\draw [line width=3]  [dash pattern={on 7.88pt off 4.5pt}]  (38.25,31.19) -- (38.25,74.85) ;
\draw [line width=3]  [dash pattern={on 7.88pt off 4.5pt}]  (59.5,31.39) -- (59.5,75.04) ;
\draw [line width=0.75]  [dash pattern={on 0.84pt off 2.51pt}]  (43.98,55.64) -- (54.4,55.77) ;
\draw [line width=3]  [dash pattern={on 7.88pt off 4.5pt}]  (110.5,74.85) -- (110.5,118.51) ;
\draw [line width=3]  [dash pattern={on 7.88pt off 4.5pt}]  (119,74.98) -- (119,118.64) ;
\draw [line width=3]  [dash pattern={on 7.88pt off 4.5pt}]  (140.25,75.04) -- (140.25,118.7) ;
\draw [line width=0.75]  [dash pattern={on 0.84pt off 2.51pt}]  (125.12,99.77) -- (135.53,99.9) ;
\draw [line width=3]    (228,118.51) -- (398,118.51) ;
\draw [shift={(322.2,118.51)}, rotate = 180] [fill={rgb, 255:red, 0; green, 0; blue, 0 }  ][line width=0.08]  [draw opacity=0] (13.57,-6.52) -- (0,0) -- (13.57,6.52) -- cycle    ;
\draw [line width=3]    (228,74.85) -- (398,74.85) ;
\draw [shift={(322.2,74.85)}, rotate = 180] [fill={rgb, 255:red, 0; green, 0; blue, 0 }  ][line width=0.08]  [draw opacity=0] (13.57,-6.52) -- (0,0) -- (13.57,6.52) -- cycle    ;
\draw [line width=3]    (228,31) -- (398,31) ;
\draw [shift={(322.2,31)}, rotate = 180] [fill={rgb, 255:red, 0; green, 0; blue, 0 }  ][line width=0.08]  [draw opacity=0] (13.57,-6.52) -- (0,0) -- (13.57,6.52) -- cycle    ;
\draw [line width=3]  [dash pattern={on 7.88pt off 4.5pt}]  (257.75,75.19) -- (257.75,118.85) ;
\draw [line width=3]  [dash pattern={on 7.88pt off 4.5pt}]  (266.25,75.19) -- (266.25,118.85) ;
\draw [line width=3]  [dash pattern={on 7.88pt off 4.5pt}]  (287.5,75.39) -- (287.5,119.04) ;
\draw [line width=0.75]  [dash pattern={on 0.84pt off 2.51pt}]  (271.98,99.64) -- (282.4,99.77) ;
\draw [line width=3]  [dash pattern={on 7.88pt off 4.5pt}]  (338.5,30.85) -- (338.5,74.51) ;
\draw [line width=3]  [dash pattern={on 7.88pt off 4.5pt}]  (347,30.98) -- (347,74.64) ;
\draw [line width=3]  [dash pattern={on 7.88pt off 4.5pt}]  (368.25,31.04) -- (368.25,74.7) ;
\draw [line width=0.75]  [dash pattern={on 0.84pt off 2.51pt}]  (353.12,55.77) -- (363.53,55.9) ;
\draw  [line width=1.5]  (191,75.38) -- (209,75.38)(200,67) -- (200,83.75) ;

\end{tikzpicture}
    \caption{ Resummation over two sets of non-crossing ladder diagrams gives the $\mathcal{O}(N^-2)$ contribution to the six-point function of fermions, these diagrams give the OPE coefficients of the integer-weight operators.} 
    \label{fig:sixpII}
\end{figure}
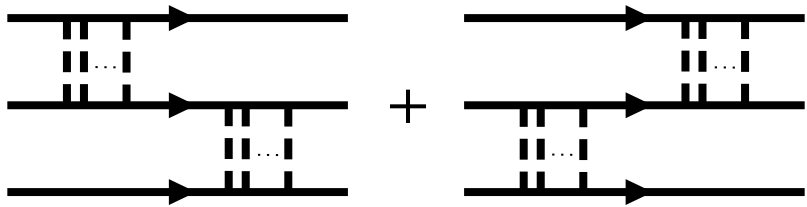
These terms give the leading order contribution to the connected six point function written in terms of the propagator $G(\vartheta)$  , 
\begin{equation}\label{eq:ferm6pt}
\begin{split}
&\frac{\braket{\bar{\psi}_i(\vartheta_1)\psi_i(\vartheta_2) \bar{\psi}_j(\vartheta_3)\psi_j(\vartheta_4)\bar{\psi}_k(\vartheta_5)\psi_k(\vartheta_6)}_C}{N^3} = \\&\hspace{2cm} \frac{3}{N^2} \left(G(\vartheta_1-\vartheta_6)G(\vartheta_3-\vartheta_2)G(\vartheta_5-\vartheta_4)+G(\vartheta_1-\vartheta_4)G(\vartheta_3-\vartheta_6)G(\vartheta_5-\vartheta_2)\right) + \mathcal{O}(N^{-3})~. 
\end{split}
\end{equation}
\subsubsection{Three-point function of operators}
We can represent the operators of the tower in terms of the original field operators of the SYK, the fermions \cite{Gross:2017aos, Gross:2017hcz, Maldacena:2016hyu}. These are primary, gaining a dimension of $\frac{1}{2}$ in the case of $q = 2$. The operators are  
\begin{equation}
    \mathcal{O}_n (\vartheta)= \frac{i}{\sqrt{N}}\sum_{r=0}^n d_{nr} \partial^r_\vartheta\bar{\psi}_i(\vartheta)\partial_\vartheta^{n-r}\psi_i(\vartheta)~, 
\end{equation}
where the $d_{nr}$ are chosen to ensure the operator is a primary, 
\begin{equation}
    d_{nr} = \frac{ \pi  J (-1)^r(-n)_r{}^2}{\sqrt{\Gamma (2 n+1)}
   \Gamma (r+1)^2}~. 
\end{equation} 
Here $(x)_n$ is the Pochhammer symbol.  These can be constructed from the normalisation of the two point function such that
\begin{equation}
\braket{\mathcal{O}_n(\vartheta_1)\mathcal{O}_m(\vartheta_2)} = \frac{\delta_{n,m} }{|\vartheta_1-\vartheta_2|^{2n}  }~.  
\end{equation}
The three point function can be retrieved from the six point function of the fermion operators in \cref{eq:ferm6pt}, by taking the appropriate derivatives and then the limits $ \vartheta_1 \rightarrow \vartheta_2$, $ \vartheta_3 \rightarrow \vartheta_4$ and $ \vartheta_5 \rightarrow \vartheta_6$. 
\begin{equation}
\begin{split}
&\braket{\mathcal{O}_n(\vartheta_2)\mathcal{O}_m(\vartheta_4)\mathcal{O}_k(\vartheta_6)}
=\\&-i\lim_{\vartheta_1 \rightarrow \vartheta_2}\lim_{\vartheta_3 \rightarrow \vartheta_4}\lim_{\vartheta_5 \rightarrow \vartheta_6}\sum_{r_{1}=0}^n\sum_{r_{2}=0}^m\sum_{r_{3}=0}^k d_{nr_1}d_{mr_2}d_{kr_3}\partial^{r_1}_{\vartheta_1}\partial_{\vartheta_2}^{n-r_1}\partial^{r_2}_{\vartheta_3}\partial_{\vartheta_4}^{m-r_2}\partial^{r_3}_{\vartheta_5}\partial_{\vartheta_6}^{k-r_3}\mathcal{F}^6_C( \vartheta_1,\vartheta_2,\vartheta_3,\vartheta_4,\vartheta_5,\vartheta_6)~. 
\end{split}
\end{equation}
Case by case the OPE coefficients can be constructed, 
\begin{equation}
    c_{nmk} = -i\sum_{r_{1}=0}^n\sum_{r_{2}=0}^m\sum_{r_{3}=0}^k d_{nr_1}d_{mr_2}d_{kr_3}\left(\frac{C_{r_1r_2r_3}\vartheta_{24}^{m+2r_1}\vartheta_{46}^{k+2r_2}\vartheta_{62}^{n+2r_3}+\tilde{C}_{r_1 r_2 r_3}\vartheta_{24}^{n+2r_2}\vartheta_{46}^{m+2r_3}\vartheta_{62}^{k+2r_1}}{\vartheta_{24}^{k+r_1+r_2}\vartheta_{46}^{n+r_3+r_2}\vartheta_{62}^{m+r_1+r_3}}\right)~.
\end{equation}
where 
\begin{align}
    C_{r_1r_2r_3}&=(-1)^{1+r_1+r_2+r_3+m+k+n}(k-r_3+r_1)!(n-r_1+r_2)!(m-r_2+r_3)!~,\\
    \tilde{C}_{r_1r_2r_3}&=(-1)^{r_1+r_2+r_3}(k-r_3+r_2)!(n-r_1+r_3)!(m-r_2+r_1)!~.
\end{align}
After resummation, these are independent of the coordinates. OPE coefficients $c_{1mk}$ with $1<m,k<5$ are displayed in \cref{tab:OPES}. 
\begin{table}
\centering
\begin{tabular}{|>{$}c<{$}||>{$}c<{$}|>{$}c<{$}|>{$}c<{$}|>{$}c<{$}|>{$}c<{$}|}
    \hline c_{1mk}& k=1 & 2 & 3 & 4 & 5 \\
    \hline\hline m=1& 0 & 3 i \sqrt{2} & 0 & \frac{3 i}{\sqrt{5}} & 0 \\
    \hline 2&3 i \sqrt{2} & 0 & 3 i \sqrt{3} & 0 & 3 i \sqrt{\frac{5}{7}} \\
    \hline 3&0 & 3 i \sqrt{3} & 0 & i \sqrt{30} & 0 \\
    \hline 4&\frac{3 i}{\sqrt{5}} & 0 & i \sqrt{30} & 0 & 3 i \sqrt{\frac{7}{2}} \\
    \hline 5&0 & 3 i \sqrt{\frac{5}{7}} & 0 & 3 i \sqrt{\frac{7}{2}} & 0 \\
    \hline
\end{tabular}
\caption{OPE coefficients $c_{1mk}$ for integer-weight operators of the complex $q=2$ SYK model for some small values of weights with $1<m,k<5$.}
\label{tab:OPES}
\end{table}
\section{Outlook\label{sec:out}}
In this note we have considered the large $N$ behaviour of the complex $q=2$ SYK model near the IR fixed point and found several differences from the  general $q>2$ case studied more regularly. As a result of its almost Gaussian structure, the model is solvable and contains a tower of integer-weight conformal operators for which we have calculated the OPE coefficients. We have therefore collected the CFT data of the single trace sector of the complex $q=2$ SYK model in preparation for further work on the bulk theory. 
\subsection{Toward a bulk description}
We have found that the model contains an infinite set of integer-weight operators, and unlike the general $q$ model there is no infinite $h=1,2$ contribution to the four-point function of fermions. This would seem to imply the model has a symmetry breaking structure distinct from the general case, in which the Schwarzian soft sector generated by the $h=2$ mode dominates the low energy behaviour of the model. Instead, we might imagine the deformation of the $q=2$ theory away from the strict IR limit contains contributions from the entire tower of operators. 

This seems to rhyme with analysis of the $SL(N, \mathbb{R})$ BF theory described in (A)dS in \cite{Anninos:2023lin,Alkalaev:2014qpa, Alkalaev:2019xuv, Alkalaev:2020kut} and particularly the higher spin analogues to the Schwarzian described in \cite{Gonzalez:2018enk}. As a result we can set up a potential holographic match between this very simple theory to a bulk theory in AdS and hopefully a development to the de Sitter case, as first conjectured in \cite{Anninos:2023lin}. There are some subtle differences between these two cases and more work is required to match the results contained in this note to new calculations in the bulk. These developments would construct a low dimensional analogue to the four-dimensional higher spin model of \cite{Anninos:2011ui,Anninos:2017eib}.
\subsection{Thermodynamics of \texorpdfstring{$q=2$}{q=2} SYK }
In future work on the $q=2$ model a thorough analysis of the thermodynamics of the disorder averaged model might provide further evidence of the emergent structure. In particular the (annealed) average free energy is calculated  at temperature $\frac{1}{\beta}$ in appendix A of \cite{Anninos:2016szt} at large $N$. As described in \cite{Gross:2016kjj,brezin1978planar} in this limit the spectral function controlling the distribution of eigenvalues of the coupling $i J_{ij}$ approximates the Wigner semicircle. This allows an analytic calculation of the annealed thermodynamic quantities in various thermodynamic limits. The calculation in the low temperature $\beta \rightarrow \infty$ limit gives a linear specific heat, and as  the free energy tends to $0$, this result agrees with the calculation of low temperature entropy for the real $q=2$ model in equation (D.5) of \cite{Anninos:2024wpy}, calculated from the master fields in  \cref{def:Effact}
\begin{equation}
    \lim_{\beta\rightarrow \infty}\braket{S[\beta]}_J = \lim_{\beta\rightarrow \infty}\braket{C[\beta]}_J = \frac{\pi N }{6 J \beta} +\mathcal{O}(N^2)~.
    \label{eq:entr}
\end{equation}
As described in \cite{Anninos:2024wpy} the entropy of the $q=2$ model is $0$ in the large $N$ limit at zero temperature in a significant departure from the general case. In \cref{App:thermo} the matching quenched disorder quantities are calculated and presented for a finite number of realisations of the real $q=2$ model using excellent freely available code \cite{Shaposnik2016}. These results were then extrapolated to large $N$ using the same procedure as in \cite{Cotler:2016fpe} to show the expected zero low temperature and linear behaviour of the entropy for the real model in \cref{fig:Entropy}. We expect the complex model to retain many of the same features, particularly as the spectral function controlling the thermodynamics converges for large $N$ \cite{Gross:2016kjj}. 

The linear entropy behavior \cref{eq:entr} can be derived from the Schwarzian action in the $q=2$ case, much like the general case. This would seem to disagree with the more democratic soft sector of the theory suggested by the absence of the double pole in the four-point function. However, it is consistent with our analysis if it can be shown that these other soft sectors contribute at higher order in $\frac{1}{\beta}$. Indeed the generality of the Schwarzian contribution for any $q$ SYK is commented on in
\cite{Mertens:2018fds}, in which a link is made to the $SL(2, \mathbb{R})$ BF theory, a subsector of the $SL(N,\mathbb{R})$ model of \cite{Alkalaev:2013fsa, Anninos:2023lin}. A more thorough analysis is required and it would seem that the numerical results presented in \cref{App:thermo} are a good starting point for this work. 
\appendix
\section*{Acknowledgements}
The author would like to thank Dionysios Anninos, Tarek Anous, Max Downing, Damian Galante, Vasileios Letsios, Alan Rios Fukelman, Vladimir Schaub, Sameer Sheorey and Andy Svesko for discussion and guidance and particularly Sameer Sheorey for help with the numerical aspects. The author would also like to extend their gratitude to the organisers and participants of the 5th Mons workshop on Higher Spin Gauge Theories for stimulating discussion and in particular Xavier Bekaert in this regard. B.P. is funded by the STFC under grant number ST/V506771/1. 
\section{Numerical analysis of Majorana \texorpdfstring{$q=2$}{q=2} SYK}\label{App:thermo}
The general $q$ SYK model has specific thermodynamic properties arising because of the presence of the $h = 2$ mode  and resultant domination of the low energy dynamics by the Schwarzian contribution as described in \cite{Maldacena:2016hyu,Cotler:2016fpe}. In this appendix we use publicly available code \cite{Shaposnik2016} to repeat the analysis of this work in the real $q = 2$ case. We work with Majorana fermions in this appendix as the thermodynamic properties in the limits discussed in the text are the same as the complex model and as the Majorana case enjoys a simpler implementation. The Hamiltonian of the model is 
\begin{equation}
    H_{SYK} = i \sum_{i\leq j}J_{ij} \psi_i \psi_j~.
\end{equation}
$J_{ij}$ is once again drawn from a real Gaussian distribution with variance $\braket{J_{ij}^2} = J^2/N$. We directly calculated the spectrum of energy eigenvalues of this Hamiltonian for $J=1$ and even $N$ values between $N=10$ and $N=30$ for $2^{\frac{36-N}{2}}$ realisations of the model. As the convergence of the density of states is independent of the number of realisations, just as in the generic $q$ case there is a ``self averaging'' property of the model which allows the scaling of the sample size with the size of the Hilbert space $2^{\frac{N}{2}}$. 
\subsection{Density of states}
\begin{figure}[ht]
    \centering
    \includegraphics[width=0.75\linewidth]{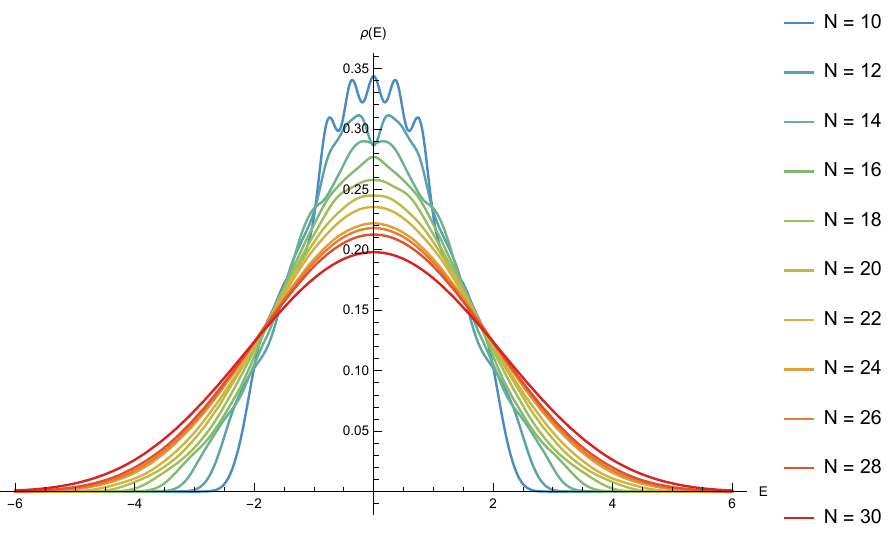}
    \caption{Normalised disorder average density of states $\rho(E)$  plotted as a smooth histogram against energy for the $J=1$, $q=2$ SYK model with increasing values of $N$. In each instance  $2^{\frac{36-N}{2}}$ realisations of the model were averaged. The density of states tends to zero smoothly with a ``tail'' indicative of a solvable theory, unlike the semi-circle $\rho(E)$ approached in the large-$N$ limit for the $q>2$ model.   }
    \label{fig:density}
\end{figure}

At finite $N$ an individual realisation of the $q=2$ model is a model of free fermions with random masses. It is possible to numerically evaluate the partition function of a large number of realisations and perform the disorder average directly for low values of N. In the $q=2$ case the density of states is fundamentally different  at all $N$ to that of larger finite values of $q>2$. \cref{fig:density} displays the normalised density of states for increasing even $N$ over $2^{\tfrac{36-N}{2}}$. The most important aspect of this difference is the smooth  decrease in the density of states at small energies. Each realisation has a single lowest energy state given by the creation of the lowest random mass particle. This results in lengthening ``tails'' of low energy, low density states, unlike the $q > 2$ case displayed in figure 15 of \cite{Cotler:2016fpe}. Much like the result for higher $q$, low $N$ realisations of the model have oscillations in the density of states caused by level repulsion. These are smoothed out as the eigenvalue spectrum of the coupling (the spectral function denoted by $\rho(\lambda)$  in \cite{Gross:2016kjj}, note this is not the density of states discussed here) approaches the Wigner semicircle \cite{Anninos:2016szt,brezin1978planar}. 
\subsection{Entropy}
\begin{figure}[ht]
    \centering
    \includegraphics[width=0.75\linewidth]{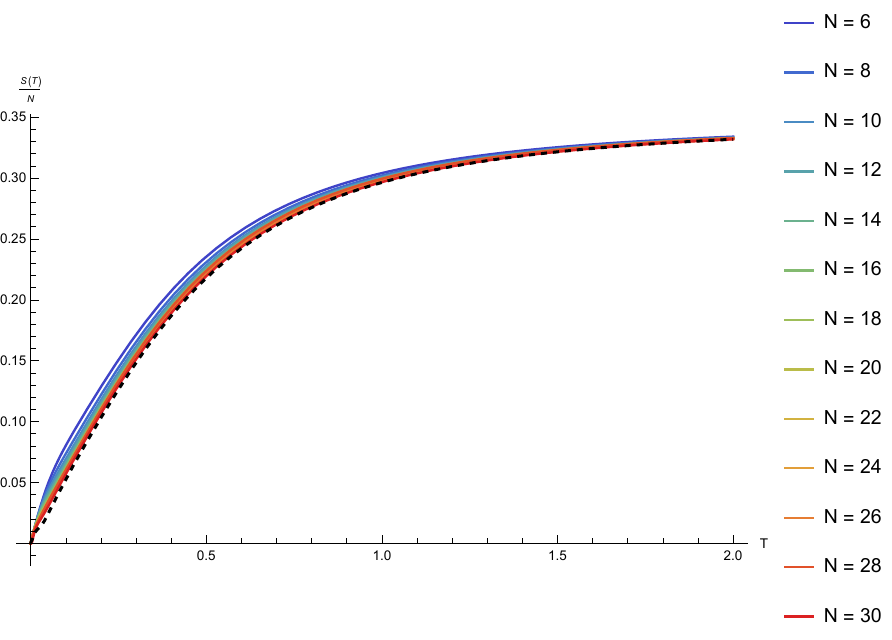}
    \caption{Entropy with respect to temperature $\frac{S(T)}{N}$ plotted from the disorder average Majorana SYK model with $q=2$ and $J=1$ for various $N$.  In each instance  $2^{\frac{36-N}{2}}$ realisations of the model were averaged as described above. Fitted with a quadratic polynomial in $N^{-1}$, the extrapolated large $N$ result is plotted as a dashed line.  }
    \label{fig:Entropy}
\end{figure}

\begin{figure}[!ht]
    \centering
    \includegraphics[width=0.75\linewidth]{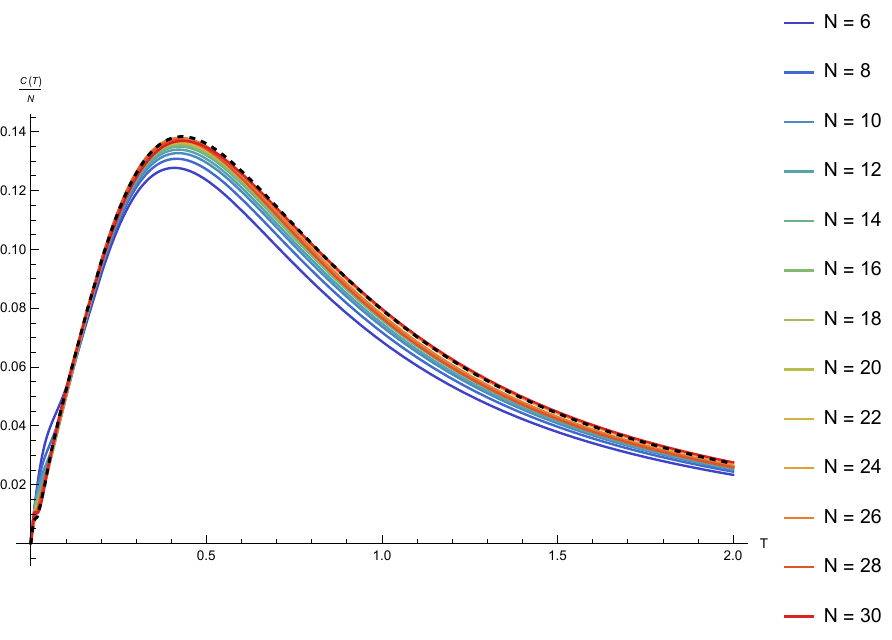}
    \caption{Specific heat with respect to temperature $\frac{C(T)}{N}$ plotted from the disorder average Majorana SYK model with $q=2$ and $J=1$ for various $N$.  In each instance  $2^{\frac{36-N}{2}}$ realisations of the model were averaged as described above. Fitted with a quadratic polynomial in $N^{-1}$, the extrapolated large $N$ result is plotted as a dashed line.  }
    \label{fig:specific}
\end{figure}

The tails of the distribution of states cause the low temperature entropy to be fixed at zero. Using the above data we calculated the disorder averaged entropy as a function of temperature for each N. We then used a large $N$ fitting adapted from that in \cite{Cotler:2016fpe}. Extrapolating from a fit of this data  at each $T$ to the polynomial 
\begin{equation}
    a(T) + \frac{b(T)}{N} + \frac{c(T)}{N^2}~, 
\end{equation}
we took the value of a(T) as the large $N$ approximation to the curve. This is displayed in \cref{fig:Entropy} as the dashed plot. It is clear that in the $q=2$ case the low temperature entropy converges on 
\begin{equation}
    \lim_{\beta \rightarrow \infty}\lim_{N \rightarrow \infty} S \approx 0~.
\end{equation}
This is consistent with the analytic expectation of the entropy in this limit (to this order) derived in the appendix of \cite{Anninos:2022qgy} for the complex model. This graph should be compared to fig. 6 of \cite{Cotler:2016fpe}, in which the same extrapolation was performed for the $q=4$ Majorana model. A finite, non-zero, low temperature, large $N$ entropy was observed in this case, consistent with
\begin{equation}
    \lim_{\beta \rightarrow \infty}\lim_{N \rightarrow \infty} S_{q>2} \approx N~.
\end{equation}
At low temperature the model exhibits a linear entropy and a linear specific heat with respect to time this is shown in \cref{fig:specific} using the same extrapolation as for the entropy. It is noted here that the large $N$ suppressed $\log(T)$ contribution to the entropy discussed in \cite{Maldacena:2016hyu, Cotler:2016fpe} was not observed in the $q=2$ case using this numerical procedure. 
\subsection{Spectral Form Factor}
\begin{figure}
    \centering
    \includegraphics[width=0.75\linewidth]{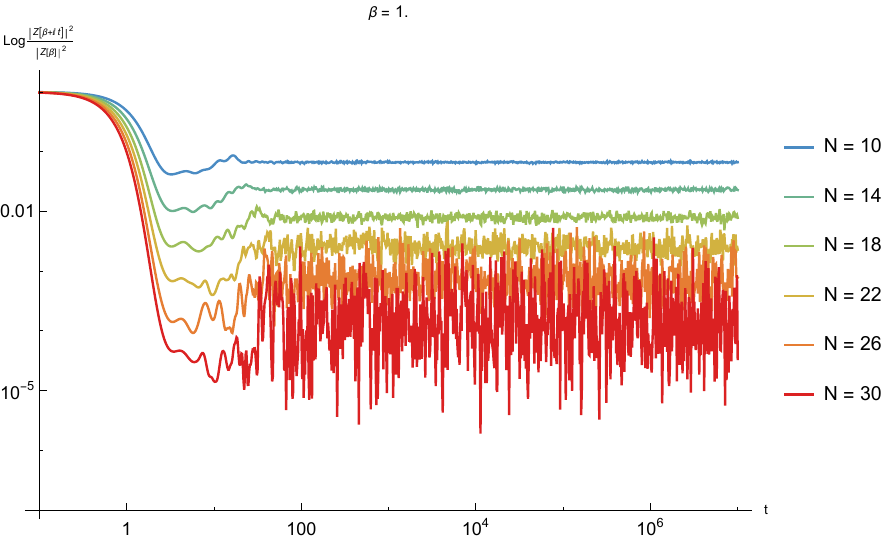}
    \caption{The spectral form factor $\log\left(\frac{Z(\beta+i t)}{Z(\beta)}\right)$ plotted w.r.t $t$, with $\beta =1$ for the disorder averaged Majorana $q=2$ SYK model with $J=1$  and various $N$ with $2^{\frac{36-N}{2}}$ realisations. }
    \label{fig:SFFGen}
\end{figure}

The spectral form factor of the $q=2$ model has been discussed in detail in appendix E of \cite{Cotler:2016fpe}. It is calculated analytically and a comment is made, that while the model is not chaotic there is evidence of a ``mini ramp'' and ``mini plateau'' in the squared disorder averaged correlation function. We calculated the disorder average square partition function and do indeed observe a slow increase after a minimum, with a ``plateau'' at $t\sim N$. However, the increasing phase is not an obviously linear ``ramp''. With increasing $N$ it also seems that this squared quantity has an increasing amount of noise around the ``plateau''. The spectral form factor for the real Majorana $q=2$ SYK model is calculated for $N = 10,16,20,24,30$ in \cref{fig:SFFGen} in agreement with \cite{Cotler:2016fpe}. 
\section{Eigenvalues of the ladder generating kernel}\label{app:evals}
We  make use of a result from \cite{Bulycheva:2017uqj}, that the eigenvalues of the action of the kernel on one of the eigenfunctions given in \cref{app:eigen} are independent of $h$. In this appendix we show this. In the CFT limit the quadratic Casimir of the conformal algebra in the space of bilinear functions operates as 
\begin{equation}
    \mathcal{C}_{12} = 2 \Delta( \Delta -1)-\hat{K}_1\hat{P}_2-\hat{P}_1\hat{K}_2\ + 2 \hat{D}_1\hat{D}_2~,
\end{equation}
where 
\begin{align}
    \hat{D}_i &=-\vartheta_i \frac{\partial}{\partial\vartheta_i}- \Delta~,\\
    \hat{P}_i &= \frac{\partial}{\partial\vartheta_i}~,\\
    \hat{K}_i &= \vartheta_i^2 \frac{\partial}{\partial \vartheta_i} + 2 \vartheta_i \Delta~.
\end{align}
The operation of integrating against the kernel commutes with the action of the Casimir \cite{Maldacena:2016hyu}. In fact as we are now considering a CFT all correlation functions are eigenfunctions of the Casimir, this includes the three point functions of two fermions (weight $ \tfrac{1}{2}$) with  another operator of weight $h$   
\begin{equation}\label{eq:splitparts}
    \braket{ \overline{\psi}_i(\vartheta_1)\psi_i(\vartheta_2)\mathcal{O}_h(\vartheta_0)} =  \frac{\text{sgn}(\vartheta _1-\vartheta_2)-i \text{sgn}(\vartheta_0-\vartheta_1)\text{sgn}(\vartheta_0- \vartheta_2)}{|\vartheta_1-\vartheta_0|^h|\vartheta_2-\vartheta_0|^h|\vartheta_1-\vartheta_2|^{1-h}}
\end{equation}
We note that the symmetric and antisymmetric in $\vartheta_1 \leftrightarrow\vartheta_2$  parts of this three point function are clearly the real and imaginary pieces. These cannot, by themselves, form the basis functions for the four point function $ \Psi_h$. However, as we noted above, the kernel \cref{def:kernelac} only acts on two of the arguments of these basis four point functions. However, we may write the four point function as an integral over these three point functions,
\begin{equation}\label{eq:split}
    \Psi_h(\vartheta_1,\vartheta_2,\vartheta_3,\vartheta_4) \approx \int d\vartheta_0     \braket{ \overline{\psi}_i(\vartheta_1)\psi_i(\vartheta_2)\mathcal{O}_h(\vartheta_0)}\braket{\mathcal{O}_h(\vartheta_0)\overline{\psi}_i(\vartheta_3)\psi_i(\vartheta_4)}~.
\end{equation}
As we are only currently interested in the eigenvalue $ k(h)$ we can ignore this decomposition for now, it is used to calculate the $\Psi_h$ in \cite{Bulycheva:2017uqj}, which we have summarised in \cref{app:eigen}. The Casimir is a differential operator acting on the variables outside of this integration, so $k(h)$ can be calculated by integrating the three-point functions against the kernel
\begin{equation}
    \int d\vartheta d\vartheta' K(1,0,\vartheta,\vartheta')\braket{ \overline{\psi}_i(\vartheta)\psi_i(\vartheta')\mathcal{O}_h(\vartheta_0)}^\pm = k^\pm(h) \braket{\overline{\psi}_i(1)\psi_i(0)\mathcal{O}_h(\vartheta_0)}~,
\end{equation}
defining
\begin{align}
\braket{ \overline{\psi}_i(\vartheta)\psi_i(\vartheta')\mathcal{O}_h(\vartheta_0)}^+ &= \Re\braket{ \overline{\psi}_i(\vartheta)\psi_i(\vartheta')\mathcal{O}_h(\vartheta_0)}~,\\
\braket{ \overline{\psi}_i(\vartheta)\psi_i(\vartheta')\mathcal{O}_h(\vartheta_0)}^- &= i\Im\braket{ \overline{\psi}_i(\vartheta)\psi_i(\vartheta')\mathcal{O}_h(\vartheta_0)}~.    
\end{align}
Using the limit $\vartheta_0 \rightarrow \infty$  and \cref{def:kern}
\begin{align}
   k^+(h) &=\pi^{-2}\int d\vartheta d\vartheta' \frac{\text{sgn}(\vartheta-1)\text{sgn}(\vartheta')\text{sgn}(\vartheta-\vartheta')}{|\vartheta-1||\vartheta'||\vartheta-\vartheta'|^{1-h}}\\
   k^-(h) &=\pi^{-2}\int d\vartheta d\vartheta' \frac{\text{sgn}(\vartheta-1)\text{sgn}(\vartheta')}{|\vartheta-1||\vartheta'||\vartheta-\vartheta'|^{1-h}}
\end{align}
Using 
\begin{equation}
    \frac{1}{\vartheta} = i \pi\int \frac{d \omega}{2\pi} \text{sgn}(\omega)e^{-i \omega \vartheta} ~,
\end{equation}
we retrieve,
\begin{equation}
    k^\pm(h)  =-\int d\omega f^\pm (\omega) e^{-i \omega} \equiv -1~,
\end{equation}
for all values of $h$. Here $f^\pm$ is the frequency space representation of the remaining function in each of the integrals above. We have made use of the fact here that both of the remaining function evaluate to $ 1$ at $\vartheta-\vartheta' = 1$. 
\section{Eigenfunctions of the one-dimensional \texorpdfstring{$SL(2,\mathbb{R})$}{SL(2,R)} Casimir }\label{app:eigen}
Our analysis requires a complete set of complex one-dimensional eigen-functions $ \tilde{\Psi}_h$ of the $SL(2, \mathbb{R})$ Casimir. 
\begin{equation}
    \mathcal{C} = x^2(1-x)\partial^2_x-x^2\partial_x~.
\end{equation}
These were calculated in \cite{Bulycheva:2017uqj}, for convenience a short summary is included here. The appropriate functions are found by solving the eigenproblem 
\begin{equation}\label{eq:cas}
    \mathcal{C} \tilde{\Psi}_h(x) = h(h-1)\tilde{\Psi}_h(x) ~,
\end{equation}
and looking  for normalisable functions under the inner product 
\begin{equation}
    \braket{f,g} = \int_{-\infty}^{\infty} \frac{dx}{x^2} \overline{f}g ~.
\end{equation}
We are interested in two sets of eigen-functions, those even and odd under the transformation $x \rightarrow \frac{x}{x-1}$, we label these 
\begin{equation}\label{def:evenodd}
    \tilde{\Psi}^+_h\left(\frac{x}{x-1}\right) = \tilde{\Psi}^+_h(x)~, \hspace{2cm} \tilde{\Psi}^-_h\left(\frac{x}{x-1}\right) = -\tilde{\Psi}^-_h(x)~.
\end{equation}
The differential equation \cref{eq:cas} has solutions 
\begin{equation}\label{def:cassoln}
    F_h = \frac{ \Gamma(h)^2}{\Gamma{(2h)}} x^h \, _2F_1(h,h;2 h;x)
\end{equation}
the other solution can be reached using the symmetry of the differential equation under $h \rightarrow 1-h$. As there is a branch cut some care needs to be taken to ensure the solutions combine with the correct behaviour under the transformation \cref{def:evenodd}, the results are 
\begin{equation}
\tilde{\Psi}_h^+(x)  = \begin{cases}
    \frac{2}{\cos\pi h} \left(\cos^2 \frac{\pi h }{2}F_h(x)-\sin^2 \frac{\pi h }{2}F_{1-h}(x) \right) \hspace{1cm} x<1~,\\\frac{2}{\sqrt{\pi}}\Gamma\left(\frac{h}{2}\right)\Gamma\left(\frac{1-h}{2}\right) \, _2F_1\left(\frac{h}{2},\frac{1-h}{2};\frac{1}{2};\frac{(2-x)^2}{x^2}\right)\hspace{1cm} x>1~,
\end{cases}
\end{equation}
and
\begin{equation}
\tilde{\Psi}_h^-(x)  = \begin{cases}
    \frac{2}{\cos\pi h} \left(\sin^2 \frac{\pi h }{2}F_h(x)-\cos^2 \frac{\pi h }{2}F_{1-h}(x) \right) \hspace{1cm} x<1~,\\\frac{4}{\sqrt{\pi}}\left(\frac{2-x}{x}\right)\Gamma\left(1-\frac{h}{2}\right)\Gamma\left(\frac{1+h}{2}\right) \, _2F_1\left(1-\frac{h}{2},\frac{1+h}{2};\frac{3}{2};\frac{(2-x)^2}{x^2}\right)\hspace{1cm} x>1~.
\end{cases}
\end{equation}
Which are normalisable up to boundary terms in the cases $ h \in \mathbb{Z}$ and $ h = \frac{1}{2} + i s$,  where $ s \in \mathbb{R}$. The inner products are respectively 
\begin{align}
    \braket{\tilde{\Psi}^+_h(x),\tilde{\Psi}^+_{h'}(x)} = \braket{\tilde{\Psi}^-_h(x),\tilde{\Psi}^-_{h'}(x)} &= \frac{2\pi^2 \delta_{h,h'}}{|h-\tfrac{1}{2}|}&h&\in\mathbb{Z}~,\\
    \braket{\tilde{\Psi}^+_s(x),\tilde{\Psi}^+_{s'}(x)} = \braket{\tilde{\Psi}^-_s(x),\tilde{\Psi}^-_{s'}(x)} &= \frac{4\pi^2 \coth \pi s}{s}\delta (s-s')&h &= \frac{1}{2} + i s, \\
    \braket{\tilde{\Psi}^+_h(x),\tilde{\Psi}^-_{h'}(x)} &=0~.
\end{align}
Throughout the above we make use of the symmetry under $ h \rightarrow 1-h$ to limit the ranges of the weights to $s >0$ and $h \in \mathbb{Z}^+$. 

As a note on the group theory, we might expect a third case in which there are normalisable wavefunctions for the above Casimir, with real weights in the complementary series $ 0< h<\frac{1}{2}$, as these weights also provide admissible unitary  representations of $SL(2, \mathbb{R})$. However these are not required to form a complete basis of functions on $\mathbb{R}$  with this inner product \cite{Maldacena:2016hyu}. 
\bibliographystyle{utphys.bst}
\bibliography{references}

\providecommand{\href}[2]{#2}\begingroup\raggedright\begin{thebibliography}{10}

\bibitem{Anninos:2023lin}
D.~Anninos, T.~Anous, B.~Pethybridge, and G.~\c{S}eng\"or, ``{The Discreet
  Charm of the Discrete Series in DS$_2$},''
  \href{http://arxiv.org/abs/2307.15832}{{\ttfamily arXiv:2307.15832
  [hep-th]}}.

\bibitem{Sachdev:1993ppp}
S.~Sachdev and J.~Ye, ``Gapless spin-fluid ground state in a random quantum
  heisenberg magnet,''
  \href{http://dx.doi.org/10.1103/PhysRevLett.70.3339}{{\em Physical Review
  Letters} {\bfseries 70} no.~21, (May, 1993) 3339--3342},
  \href{http://arxiv.org/abs/cond-mat/9212030}{{\ttfamily
  arXiv:cond-mat/9212030}}.

\bibitem{Kitaev:2015}
A.~Kitaev, ``A simple model of quantum holography.'' Talks at kavli institute
  for theoretical physics, santa barbara u.s.a., 2015.
\newblock \url{https://online.kitp.ucsb.edu/online/entangled15/kitaev/}
  \url{https://online.kitp.ucsb.edu/online/entangled15/kitaev2/}.

\bibitem{Maldacena:2016hyu}
J.~Maldacena and D.~Stanford, ``{Remarks on the Sachdev-Ye-Kitaev model},''
  \href{http://dx.doi.org/10.1103/PhysRevD.94.106002}{{\em Phys. Rev. D}
  {\bfseries 94} no.~10, (2016) 106002},
  \href{http://arxiv.org/abs/1604.07818}{{\ttfamily arXiv:1604.07818
  [hep-th]}}.

\bibitem{Gross:2017hcz}
D.~J. Gross and V.~Rosenhaus, ``{The Bulk Dual of SYK: Cubic Couplings},''
  \href{http://dx.doi.org/10.1007/JHEP05(2017)092}{{\em JHEP} {\bfseries 05}
  (2017) 092}, \href{http://arxiv.org/abs/1702.08016}{{\ttfamily
  arXiv:1702.08016 [hep-th]}}.

\bibitem{Bulycheva:2017uqj}
K.~Bulycheva, ``{A note on the SYK model with complex fermions},''
  \href{http://dx.doi.org/10.1007/JHEP12(2017)069}{{\em JHEP} {\bfseries 12}
  (2017) 069}, \href{http://arxiv.org/abs/1706.07411}{{\ttfamily
  arXiv:1706.07411 [hep-th]}}.

\bibitem{Gu:2019jub}
Y.~Gu, A.~Kitaev, S.~Sachdev, and G.~Tarnopolsky, ``{Notes on the complex
  Sachdev-Ye-Kitaev model},''
  \href{http://dx.doi.org/10.1007/JHEP02(2020)157}{{\em JHEP} {\bfseries 02}
  (2020) 157}, \href{http://arxiv.org/abs/1910.14099}{{\ttfamily
  arXiv:1910.14099 [hep-th]}}.

\bibitem{Jackiw:1984je}
R.~Jackiw, ``{Lower Dimensional Gravity},''
  \href{http://dx.doi.org/10.1016/0550-3213(85)90448-1}{{\em Nucl. Phys. B}
  {\bfseries 252} (1985) 343--356}.

\bibitem{Teitelboim:1983ux}
C.~Teitelboim, ``{Gravitation and Hamiltonian Structure in Two Space-Time
  Dimensions},'' \href{http://dx.doi.org/10.1016/0370-2693(83)90012-6}{{\em
  Phys. Lett. B} {\bfseries 126} (1983) 41--45}.

\bibitem{Almheiri:2014cka}
A.~Almheiri and J.~Polchinski, ``{Models of AdS$_{2}$ backreaction and
  holography},'' \href{http://dx.doi.org/10.1007/JHEP11(2015)014}{{\em JHEP}
  {\bfseries 11} (2015) 014}, \href{http://arxiv.org/abs/1402.6334}{{\ttfamily
  arXiv:1402.6334 [hep-th]}}.

\bibitem{Maldacena:2016upp}
J.~Maldacena, D.~Stanford, and Z.~Yang, ``{Conformal symmetry and its breaking
  in two dimensional Nearly Anti-de-Sitter space},''
  \href{http://dx.doi.org/10.1093/ptep/ptw124}{{\em PTEP} {\bfseries 2016}
  no.~12, (2016) 12C104}, \href{http://arxiv.org/abs/1606.01857}{{\ttfamily
  arXiv:1606.01857 [hep-th]}}.

\bibitem{Cotler:2016fpe}
J.~S. Cotler, G.~Gur-Ari, M.~Hanada, J.~Polchinski, P.~Saad, S.~H. Shenker,
  D.~Stanford, A.~Streicher, and M.~Tezuka, ``{Black Holes and Random
  Matrices},'' \href{http://dx.doi.org/10.1007/JHEP05(2017)118}{{\em JHEP}
  {\bfseries 05} (2017) 118}, \href{http://arxiv.org/abs/1611.04650}{{\ttfamily
  arXiv:1611.04650 [hep-th]}}. [Erratum: JHEP 09, 002 (2018)].

\bibitem{Jensen:2016pah}
K.~Jensen, ``{Chaos in AdS$_2$ Holography},''
  \href{http://dx.doi.org/10.1103/PhysRevLett.117.111601}{{\em Phys. Rev.
  Lett.} {\bfseries 117} no.~11, (2016) 111601},
  \href{http://arxiv.org/abs/1605.06098}{{\ttfamily arXiv:1605.06098
  [hep-th]}}.

\bibitem{Kitaev:2017awl}
A.~Kitaev and S.~J. Suh, ``{The soft mode in the Sachdev-Ye-Kitaev model and
  its gravity dual},'' \href{http://dx.doi.org/10.1007/JHEP05(2018)183}{{\em
  JHEP} {\bfseries 05} (2018) 183},
  \href{http://arxiv.org/abs/1711.08467}{{\ttfamily arXiv:1711.08467
  [hep-th]}}.

\bibitem{Joung:2023doq}
E.~Joung, P.~Narayan, and J.~Yoon, ``{Gravitational Edge Mode in Asymptotically
  AdS$_2$: JT Gravity Revisited},''
  \href{http://arxiv.org/abs/2304.06088}{{\ttfamily arXiv:2304.06088
  [hep-th]}}.

\bibitem{Sarosi:2017ykf}
G.~S\'arosi, ``{AdS$_{2}$ holography and the SYK model},''
  \href{http://dx.doi.org/10.22323/1.323.0001}{{\em PoS} {\bfseries Modave2017}
  (2018) 001}, \href{http://arxiv.org/abs/1711.08482}{{\ttfamily
  arXiv:1711.08482 [hep-th]}}.

\bibitem{Rosenhaus:2018dtp}
V.~Rosenhaus, ``{An introduction to the SYK model},''
  \href{http://dx.doi.org/10.1088/1751-8121/ab2ce1}{{\em J. Phys. A} {\bfseries
  52} (2019) 323001}, \href{http://arxiv.org/abs/1807.03334}{{\ttfamily
  arXiv:1807.03334 [hep-th]}}.

\bibitem{Alkalaev:2013fsa}
K.~B. Alkalaev, ``{On higher spin extension of the Jackiw-Teitelboim gravity
  model},'' \href{http://dx.doi.org/10.1088/1751-8113/47/36/365401}{{\em J.
  Phys. A} {\bfseries 47} (2014) 365401},
  \href{http://arxiv.org/abs/1311.5119}{{\ttfamily arXiv:1311.5119 [hep-th]}}.

\bibitem{Grumiller:2013swa}
D.~Grumiller, M.~Leston, and D.~Vassilevich, ``{Anti-de Sitter holography for
  gravity and higher spin theories in two dimensions},''
  \href{http://dx.doi.org/10.1103/PhysRevD.89.044001}{{\em Phys. Rev. D}
  {\bfseries 89} no.~4, (2014) 044001},
  \href{http://arxiv.org/abs/1311.7413}{{\ttfamily arXiv:1311.7413 [hep-th]}}.

\bibitem{Bengtsson:1986zm}
A.~K.~H. Bengtsson and I.~Bengtsson, ``{Higher 'Spins' in One and Two
  Space-time Dimensions},''
  \href{http://dx.doi.org/10.1016/0370-2693(86)91102-0}{{\em Phys. Lett. B}
  {\bfseries 174} (1986) 294--300}.

\bibitem{Vasiliev:1995sv}
M.~A. Vasiliev, ``{Higher spin gauge interactions for matter fields in
  two-dimensions},'' \href{http://dx.doi.org/10.1016/0370-2693(95)01122-7}{{\em
  Phys. Lett. B} {\bfseries 363} (1995) 51--57},
  \href{http://arxiv.org/abs/hep-th/9511063}{{\ttfamily arXiv:hep-th/9511063}}.

\bibitem{Alkalaev:2020kut}
K.~Alkalaev and X.~Bekaert, ``{On BF-type higher-spin actions in two
  dimensions},'' \href{http://dx.doi.org/10.1007/JHEP05(2020)158}{{\em JHEP}
  {\bfseries 05} (2020) 158}, \href{http://arxiv.org/abs/2002.02387}{{\ttfamily
  arXiv:2002.02387 [hep-th]}}.

\bibitem{Gonzalez:2018enk}
H.~A. Gonz\'alez, D.~Grumiller, and J.~Salzer, ``{Towards a bulk description of
  higher spin SYK},'' \href{http://dx.doi.org/10.1007/JHEP05(2018)083}{{\em
  JHEP} {\bfseries 05} (2018) 083},
  \href{http://arxiv.org/abs/1802.01562}{{\ttfamily arXiv:1802.01562
  [hep-th]}}.

\bibitem{Alkalaev:2019xuv}
K.~Alkalaev and X.~Bekaert, ``{Towards higher-spin AdS$_2$/CFT$_1$
  holography},'' \href{http://dx.doi.org/10.1007/JHEP04(2020)206}{{\em JHEP}
  {\bfseries 04} (2020) 206}, \href{http://arxiv.org/abs/1911.13212}{{\ttfamily
  arXiv:1911.13212 [hep-th]}}.

\bibitem{Penedones:2023uqc}
J.~Penedones, K.~Salehi~Vaziri, and Z.~Sun, ``{Hilbert space of Quantum Field
  Theory in de Sitter spacetime},''
  \href{http://arxiv.org/abs/2301.04146}{{\ttfamily arXiv:2301.04146
  [hep-th]}}.

\bibitem{Baumann:2017jvh}
D.~Baumann, G.~Goon, H.~Lee, and G.~L. Pimentel, ``{Partially Massless Fields
  During Inflation},'' \href{http://dx.doi.org/10.1007/JHEP04(2018)140}{{\em
  JHEP} {\bfseries 04} (2018) 140},
  \href{http://arxiv.org/abs/1712.06624}{{\ttfamily arXiv:1712.06624
  [hep-th]}}.

\bibitem{Pethybridge:2021rwf}
B.~Pethybridge and V.~Schaub, ``{Tensors and spinors in de Sitter space},''
  \href{http://dx.doi.org/10.1007/JHEP06(2022)123}{{\em JHEP} {\bfseries 06}
  (2022) 123}, \href{http://arxiv.org/abs/2111.14899}{{\ttfamily
  arXiv:2111.14899 [hep-th]}}.

\bibitem{Schaub:2023scu}
V.~Schaub, ``{Spinors in (Anti-)de Sitter Space},''
  \href{http://dx.doi.org/10.1007/JHEP09(2023)142}{{\em JHEP} {\bfseries 09}
  (2023) 142}, \href{http://arxiv.org/abs/2302.08535}{{\ttfamily
  arXiv:2302.08535 [hep-th]}}.

\bibitem{Letsios:2023qzq}
V.~A. Letsios, ``{(Non-)unitarity of strictly and partially massless fermions
  on de Sitter space},'' \href{http://dx.doi.org/10.1007/JHEP05(2023)015}{{\em
  JHEP} {\bfseries 05} (2023) 015},
  \href{http://arxiv.org/abs/2303.00420}{{\ttfamily arXiv:2303.00420
  [hep-th]}}.

\bibitem{Letsios:2024nmf}
V.~A. Letsios, ``{(Non-)unitarity of strictly and partially massless fermions
  on de Sitter space II: an explanation based on the group-theoretic properties
  of the spin-3/2 and spin-5/2 eigenmodes},''
  \href{http://dx.doi.org/10.1088/1751-8121/ad2c27}{{\em Journal of Physics A:
  Mathematical and Theoretical} (2024) }.
  \url{http://iopscience.iop.org/article/10.1088/1751-8121/ad2c27}.

\bibitem{RiosFukelman:2023mgq}
A.~Rios~Fukelman, M.~Semp\'e, and G.~A. Silva, ``{Notes on gauge fields and
  discrete series representations in de Sitter spacetimes},''
  \href{http://dx.doi.org/10.1007/JHEP01(2024)011}{{\em JHEP} {\bfseries 01}
  (2024) 011}, \href{http://arxiv.org/abs/2310.14955}{{\ttfamily
  arXiv:2310.14955 [hep-th]}}.

\bibitem{Witten:2001kn}
E.~Witten, ``{Quantum gravity in de Sitter space},'' in {\em {Strings 2001:
  International Conference}}.
\newblock 6, 2001.
\newblock \href{http://arxiv.org/abs/hep-th/0106109}{{\ttfamily
  arXiv:hep-th/0106109}}.

\bibitem{Strominger:2001pn}
A.~Strominger, ``{The dS / CFT correspondence},''
  \href{http://dx.doi.org/10.1088/1126-6708/2001/10/034}{{\em JHEP} {\bfseries
  10} (2001) 034}, \href{http://arxiv.org/abs/hep-th/0106113}{{\ttfamily
  arXiv:hep-th/0106113}}.

\bibitem{Maldacena:2002vr}
J.~M. Maldacena, ``{Non-Gaussian features of primordial fluctuations in single
  field inflationary models},''
  \href{http://dx.doi.org/10.1088/1126-6708/2003/05/013}{{\em JHEP} {\bfseries
  05} (2003) 013}, \href{http://arxiv.org/abs/astro-ph/0210603}{{\ttfamily
  arXiv:astro-ph/0210603}}.

\bibitem{Anninos:2011ui}
D.~Anninos, T.~Hartman, and A.~Strominger, ``{Higher Spin Realization of the
  dS/CFT Correspondence},''
  \href{http://dx.doi.org/10.1088/1361-6382/34/1/015009}{{\em Class. Quant.
  Grav.} {\bfseries 34} no.~1, (2017) 015009},
  \href{http://arxiv.org/abs/1108.5735}{{\ttfamily arXiv:1108.5735 [hep-th]}}.

\bibitem{Anninos:2017eib}
D.~Anninos, F.~Denef, R.~Monten, and Z.~Sun, ``{Higher Spin de Sitter Hilbert
  Space},'' \href{http://dx.doi.org/10.1007/JHEP10(2019)071}{{\em JHEP}
  {\bfseries 10} (2019) 071}, \href{http://arxiv.org/abs/1711.10037}{{\ttfamily
  arXiv:1711.10037 [hep-th]}}.

\bibitem{Martinec:2014uva}
E.~J. Martinec and W.~E. Moore, ``{Modeling Quantum Gravity Effects in
  Inflation},'' \href{http://dx.doi.org/10.1007/JHEP07(2014)053}{{\em JHEP}
  {\bfseries 07} (2014) 053}, \href{http://arxiv.org/abs/1401.7681}{{\ttfamily
  arXiv:1401.7681 [hep-th]}}.

\bibitem{Bautista:2015wqy}
T.~Bautista and A.~Dabholkar, ``{Quantum Cosmology Near Two Dimensions},''
  \href{http://dx.doi.org/10.1103/PhysRevD.94.044017}{{\em Phys. Rev. D}
  {\bfseries 94} no.~4, (2016) 044017},
  \href{http://arxiv.org/abs/1511.07450}{{\ttfamily arXiv:1511.07450
  [hep-th]}}.

\bibitem{Bautista:2019jau}
T.~Bautista, A.~Dabholkar, and H.~Erbin, ``{Quantum Gravity from Timelike
  Liouville theory},'' \href{http://dx.doi.org/10.1007/JHEP10(2019)284}{{\em
  JHEP} {\bfseries 10} (2019) 284},
  \href{http://arxiv.org/abs/1905.12689}{{\ttfamily arXiv:1905.12689
  [hep-th]}}.

\bibitem{Anninos:2021ene}
D.~Anninos, T.~Bautista, and B.~M\"uhlmann, ``{The two-sphere partition
  function in two-dimensional quantum gravity},''
  \href{http://dx.doi.org/10.1007/JHEP09(2021)116}{{\em JHEP} {\bfseries 09}
  (2021) 116}, \href{http://arxiv.org/abs/2106.01665}{{\ttfamily
  arXiv:2106.01665 [hep-th]}}.

\bibitem{Muhlmann:2021clm}
B.~M\"uhlmann, ``{The two-sphere partition function in two-dimensional quantum
  gravity at fixed area},''
  \href{http://dx.doi.org/10.1007/JHEP09(2021)189}{{\em JHEP} {\bfseries 09}
  no.~189, (2021) 189}, \href{http://arxiv.org/abs/2106.04532}{{\ttfamily
  arXiv:2106.04532 [hep-th]}}.

\bibitem{Coleman:2021nor}
E.~Coleman, E.~A. Mazenc, V.~Shyam, E.~Silverstein, R.~M. Soni, G.~Torroba, and
  S.~Yang, ``{De Sitter microstates from T$ \overline{T} $ +
  \ensuremath{\Lambda}$_{2}$ and the Hawking-Page transition},''
  \href{http://dx.doi.org/10.1007/JHEP07(2022)140}{{\em JHEP} {\bfseries 07}
  (2022) 140}, \href{http://arxiv.org/abs/2110.14670}{{\ttfamily
  arXiv:2110.14670 [hep-th]}}.

\bibitem{Muhlmann:2022duj}
B.~M\"uhlmann, ``{The two-sphere partition function from timelike Liouville
  theory at three-loop order},''
  \href{http://dx.doi.org/10.1007/JHEP05(2022)057}{{\em JHEP} {\bfseries 05}
  (2022) 057}, \href{http://arxiv.org/abs/2202.04549}{{\ttfamily
  arXiv:2202.04549 [hep-th]}}.

\bibitem{Maldacena:2019cbz}
J.~Maldacena, G.~J. Turiaci, and Z.~Yang, ``{Two dimensional Nearly de Sitter
  gravity},'' \href{http://dx.doi.org/10.1007/JHEP01(2021)139}{{\em JHEP}
  {\bfseries 01} (2021) 139}, \href{http://arxiv.org/abs/1904.01911}{{\ttfamily
  arXiv:1904.01911 [hep-th]}}.

\bibitem{Cotler:2019nbi}
J.~Cotler, K.~Jensen, and A.~Maloney, ``{Low-dimensional de Sitter quantum
  gravity},'' \href{http://dx.doi.org/10.1007/JHEP06(2020)048}{{\em JHEP}
  {\bfseries 06} (2020) 048}, \href{http://arxiv.org/abs/1905.03780}{{\ttfamily
  arXiv:1905.03780 [hep-th]}}.

\bibitem{Svesko:2022txo}
A.~Svesko, E.~Verheijden, E.~P. Verlinde, and M.~R. Visser, ``{Quasi-local
  energy and microcanonical entropy in two-dimensional nearly de Sitter
  gravity},'' \href{http://dx.doi.org/10.1007/JHEP08(2022)075}{{\em JHEP}
  {\bfseries 08} (2022) 075}, \href{http://arxiv.org/abs/2203.00700}{{\ttfamily
  arXiv:2203.00700 [hep-th]}}.

\bibitem{Nanda:2023wne}
K.~K. Nanda, S.~K. Sake, and S.~P. Trivedi, ``{JT Gravity in de Sitter Space
  and the Problem of Time},'' \href{http://arxiv.org/abs/2307.15900}{{\ttfamily
  arXiv:2307.15900 [hep-th]}}.

\bibitem{Anninos:2022hqo}
D.~Anninos and E.~Harris, ``{Interpolating geometries and the stretched
  dS$_{2}$ horizon},'' \href{http://dx.doi.org/10.1007/JHEP11(2022)166}{{\em
  JHEP} {\bfseries 11} (2022) 166},
  \href{http://arxiv.org/abs/2209.06144}{{\ttfamily arXiv:2209.06144
  [hep-th]}}.

\bibitem{Anninos:2017hhn}
D.~Anninos and D.~M. Hofman, ``{Infrared Realization of dS$_2$ in AdS$_2$},''
  \href{http://dx.doi.org/10.1088/1361-6382/aab143}{{\em Class. Quant. Grav.}
  {\bfseries 35} no.~8, (2018) 085003},
  \href{http://arxiv.org/abs/1703.04622}{{\ttfamily arXiv:1703.04622
  [hep-th]}}.

\bibitem{Anninos:2018svg}
D.~Anninos, D.~A. Galante, and D.~M. Hofman, ``{De Sitter horizons \&
  holographic liquids},'' \href{http://dx.doi.org/10.1007/JHEP07(2019)038}{{\em
  JHEP} {\bfseries 07} (2019) 038},
  \href{http://arxiv.org/abs/1811.08153}{{\ttfamily arXiv:1811.08153
  [hep-th]}}.

\bibitem{Anninos:2022qgy}
D.~Anninos, D.~A. Galante, and S.~U. Sheorey, ``{Renormalisation group flows of
  deformed SYK models},'' \href{http://dx.doi.org/10.1007/JHEP11(2023)197}{{\em
  JHEP} {\bfseries 11} (2023) 197},
  \href{http://arxiv.org/abs/2212.04944}{{\ttfamily arXiv:2212.04944
  [hep-th]}}.

\bibitem{Anninos:2024wpy}
D.~Anninos, D.~A. Galante, and C.~Maneerat, ``{Cosmological Observatories},''
  \href{http://arxiv.org/abs/2402.04305}{{\ttfamily arXiv:2402.04305
  [hep-th]}}.

\bibitem{Susskind:2021esx}
L.~Susskind, ``{Entanglement and Chaos in De Sitter Space Holography: An SYK
  Example},'' \href{http://dx.doi.org/10.22128/jhap.2021.455.1005}{{\em JHAP}
  {\bfseries 1} no.~1, (2021) 1--22},
  \href{http://arxiv.org/abs/2109.14104}{{\ttfamily arXiv:2109.14104
  [hep-th]}}.

\bibitem{Rahman:2022jsf}
A.~A. Rahman, ``{dS JT Gravity and Double-Scaled SYK},''
  \href{http://arxiv.org/abs/2209.09997}{{\ttfamily arXiv:2209.09997
  [hep-th]}}.

\bibitem{Rahman:2024vyg}
A.~A. Rahman and L.~Susskind, ``{Infinite Temperature is Not So Infinite: The
  Many Temperatures of de Sitter Space},''
  \href{http://arxiv.org/abs/2401.08555}{{\ttfamily arXiv:2401.08555
  [hep-th]}}.

\bibitem{Narovlansky:2023lfz}
V.~Narovlansky and H.~Verlinde, ``{Double-scaled SYK and de Sitter
  Holography},'' \href{http://arxiv.org/abs/2310.16994}{{\ttfamily
  arXiv:2310.16994 [hep-th]}}.

\bibitem{A:2023psv}
S.~A, T.~Banks, and W.~Fischler, ``{Quantum theory of three-dimensional de
  Sitter space},'' \href{http://dx.doi.org/10.1103/PhysRevD.109.025011}{{\em
  Phys. Rev. D} {\bfseries 109} no.~2, (2024) 025011},
  \href{http://arxiv.org/abs/2306.05264}{{\ttfamily arXiv:2306.05264
  [hep-th]}}.

\bibitem{Verlinde:2024znh}
H.~Verlinde, ``{Double-scaled SYK, Chords and de Sitter Gravity},''
  \href{http://arxiv.org/abs/2402.00635}{{\ttfamily arXiv:2402.00635
  [hep-th]}}.

\bibitem{Afshar:2019axx}
H.~Afshar, H.~A. Gonz\'alez, D.~Grumiller, and D.~Vassilevich, ``{Flat space
  holography and the complex Sachdev-Ye-Kitaev model},''
  \href{http://dx.doi.org/10.1103/PhysRevD.101.086024}{{\em Phys. Rev. D}
  {\bfseries 101} no.~8, (2020) 086024},
  \href{http://arxiv.org/abs/1911.05739}{{\ttfamily arXiv:1911.05739
  [hep-th]}}.

\bibitem{Gross:2016kjj}
D.~J. Gross and V.~Rosenhaus, ``{A Generalization of Sachdev-Ye-Kitaev},''
  \href{http://dx.doi.org/10.1007/JHEP02(2017)093}{{\em JHEP} {\bfseries 02}
  (2017) 093}, \href{http://arxiv.org/abs/1610.01569}{{\ttfamily
  arXiv:1610.01569 [hep-th]}}.

\bibitem{Gross:2017aos}
D.~J. Gross and V.~Rosenhaus, ``{All point correlation functions in SYK},''
  \href{http://dx.doi.org/10.1007/JHEP12(2017)148}{{\em JHEP} {\bfseries 12}
  (2017) 148}, \href{http://arxiv.org/abs/1710.08113}{{\ttfamily
  arXiv:1710.08113 [hep-th]}}.

\bibitem{Alkalaev:2014qpa}
K.~B. Alkalaev, ``{Global and local properties of AdS$_{2}$ higher spin
  gravity},'' \href{http://dx.doi.org/10.1007/JHEP10(2014)122}{{\em JHEP}
  {\bfseries 10} (2014) 122}, \href{http://arxiv.org/abs/1404.5330}{{\ttfamily
  arXiv:1404.5330 [hep-th]}}.

\bibitem{Anninos:2016szt}
D.~Anninos, T.~Anous, and F.~Denef, ``{Disordered Quivers and Cold Horizons},''
  \href{http://dx.doi.org/10.1007/JHEP12(2016)071}{{\em JHEP} {\bfseries 12}
  (2016) 071}, \href{http://arxiv.org/abs/1603.00453}{{\ttfamily
  arXiv:1603.00453 [hep-th]}}.

\bibitem{brezin1978planar}
E.~Br{\'e}zin, C.~Itzykson, G.~Parisi, and J.-B. Zuber, ``Planar diagrams,''
  {\em Communications in Mathematical Physics} {\bfseries 59} (1978) 35--51.

\bibitem{Shaposnik2016}
F.~I.~S. Massolo, ``Mathematica,'' 2022.
\newblock \url{https://fidel-schaposnik.github.io/mathematica/}.

\bibitem{Mertens:2018fds}
T.~G. Mertens, ``{The Schwarzian theory \textemdash{} origins},''
  \href{http://dx.doi.org/10.1007/JHEP05(2018)036}{{\em JHEP} {\bfseries 05}
  (2018) 036}, \href{http://arxiv.org/abs/1801.09605}{{\ttfamily
  arXiv:1801.09605 [hep-th]}}.

\end{thebibliography}\endgroup
\end{document}